\documentclass[sigconf,authorversion,nonacm]{acmart}

\usepackage{amsthm}
\usepackage{amsfonts}
\usepackage{alltt}
\usepackage{subfigure}
\usepackage{hyperref}
\usepackage{bbding}
\usepackage{enumitem}
\usepackage{listings}
\usepackage{balance}

\setitemize{noitemsep,topsep=0pt,parsep=0pt,partopsep=0pt,leftmargin=*}

\newcommand\vldbpagestyle{plain}

\theoremstyle{definition}

\newtheorem{example}{Example}
\newtheorem{complex_query}{Complex~Query}

\lstdefinestyle{C++Style}{
  basicstyle=\ttfamily\footnotesize,
  columns=fullflexible,
  breakatwhitespace=false,
  breaklines=true,
  extendedchars=true,
  frame=single,
  keepspaces=true,
  keywordstyle=\color{blue},
  language=c++,
  numbers=none,
  numberstyle=\tiny\color{blue},
  rulecolor=\color{white},
  showspaces=false,
  showtabs=false,
}

\lstdefinestyle{QueryStyle}{
  basicstyle=\ttfamily,
  columns=fullflexible,
  breaklines=true,
  extendedchars=true,
  frame=single,
  keepspaces=true,
  numbersep=5pt,
  numberstyle=\tiny\color{blue},
  rulecolor=\color{white},
}

\newcommand{\gqs}{Banyan}

\newcommand{\Figure}{Figure}

\begin{document}

\title{Banyan: A Scoped Dataflow Engine for Graph Query Service}

\author{Li Su\texorpdfstring{$^{1}$}{Lg}, Xiaoming Qin\texorpdfstring{$^{1}$}{Lg},  Zichao Zhang\texorpdfstring{$^{1}$}{Lg}, Rui Yang\texorpdfstring{$^{2}$}{Lg}, Le Xu\texorpdfstring{$^{2}$}{Lg}, Indranil Gupta\texorpdfstring{$^{2}$}{Lg}, \break
Wenyuan Yu\texorpdfstring{$^{1}$}{Lg}, Kai Zeng\texorpdfstring{$^{1}$}{Lg}, Jingren Zhou\texorpdfstring{$^{1}$}{Lg}}
\affiliation{
	\smallskip \smallskip
	\institution{\LARGE \textit{\texorpdfstring{$^{1}$}{Lg}Alibaba Group, \texorpdfstring{$^{2}$}{Lg}University of Illinois at Urbana Champaign}}
	\smallskip \smallskip
	\state{\Large \textsf\texorpdfstring{$^1\{$}lisu.sl, xiaoming.qxm, houbai.zzc, wenyuan.ywy, zengkai.zk, jingren.zhou\texorpdfstring{$\}$}{Lg}@alibaba-inc.com
			\break \texorpdfstring{$^2\{$}{Lg}ry2, lexu1, indy\texorpdfstring{$\}$}{Lg}@illinois.edu
	\country{}
}}

\begin{abstract}
Graph query services (GQS) are widely used today to interactively answer graph
traversal queries on large-scale graph data.
Existing graph query engines focus largely on optimizing the latency of a single query. This ignores significant challenges posed by GQS, including fine-grained control and scheduling during query execution, as well as performance isolation and load balancing in various levels from across user to intra-query.
To tackle these control and scheduling challenges, we propose a novel \textit{scoped} dataflow for modeling graph traversal queries, which explicitly exposes concurrent execution and control of any subquery to the finest granularity. We implemented \gqs{}, an engine based on the scoped dataflow model for GQS.
\gqs{} focuses on scaling up the performance on a single machine,
and provides the ability to easily scale out.
Extensive experiments on multiple benchmarks show that \gqs{}
improves performance by up to three orders of magnitude over state-of-the-art
graph query engines, while providing performance isolation and load balancing.
\end{abstract}

\maketitle

\pagestyle{\vldbpagestyle}

\section{Introduction}\label{sec:introduction}

Graph query service (GQS) is widely used in many Internet applications
ranging from search engines, recommendation systems, to financial risk management.
The global GQS market is estimated to reach $2.9$ billion USD by 2024,
with an annual growth rate of $22.2\%$~\cite{graphdb-annual-growth}.
In these applications, data are represented as large-scale graphs
such as knowledge graphs and social networks, and the explorations on
the data are usually expressed as graph traversal queries.
GQS serves these large-scale graphs for interactive query access,
allowing many users to submit concurrent graph traversal queries
and obtain results in real-time.

Previous graph query engines~\cite{zhao2010graph, yan2016general, sakr2012g, shao2013trinity,
neo4j, janusgraph} mainly focus on optimizing the query latency from the perspective of
computation efficiency, i.e., traversing more vertices/edges in a time unit.
However, we observe that optimizing for computational efficiency \textit{alone}
is not sufficient to achieve short query latency. A multi-tenant GQS needs
to address two key goals to fulfill stringent latency requirements:
(\textbf{O1}) fine-grained control and scheduling,
as well as (\textbf{O2}) performance isolation and load balancing
during the query execution.

\begin{table*}[t]
\footnotesize
\centering
\begin{tabular}{|l|c|c|c|c|c|}
\hline
                                                   			& \textbf{Neo4j} & \textbf{JanusGraph} & \textbf{Timely} & \textbf{GAIA} & \textbf{Banyan}  \\ \hline
\textbf{O1-1: Control on Subquery Traversal}   & \XSolidBrush   & \XSolidBrush        & \XSolidBrush        & \Checkmark    & \Checkmark       \\ \hline
\textbf{O1-2: Subquery-level Scheduling Policy}    			& \XSolidBrush   & query-level only        & \XSolidBrush        & query-level only  & \Checkmark       \\ \hline
\textbf{O2: Subquery-level Isolation} 	& \XSolidBrush   & \XSolidBrush        & \XSolidBrush        & \XSolidBrush  & \Checkmark       \\ \hline
\end{tabular}
\caption{Comparing graph query engines on \textbf{O1} and \textbf{O2}.
TigerGraph~\cite{tigergraph} is not open-sourced and thus not listed.}
\label{table:systems}
\vspace{-15pt}
\end{table*}

\noindent\textbf{Fine-Grained Control and Scheduling (O1).}
A graph query usually performs many traversals starting
from different source vertices in the graph.
We observe two key techniques that can reduce query latencies,
and in turn boost the system throughput:
(\textbf{O1-1}) concurrently executing these traversals,
and controlling them at a fine granularity;
(\textbf{O1-2}) carefully choosing the traversal strategy.

\newpage
\begin{example}[A Graph Traversal Query in Gremlin]

\begin{lstlisting}[style=QueryStyle, upquote=false]
g.V(123).repeat(out(`knows'))
    .until(out(`worksAt').is(eq(`XYZ')))
    .or().loops().is(gt(5)))
 .where(in(`tweets').out(`hasTag').is(eq(`#ABC')))
 .limit(20)
\end{lstlisting}
\label{ex:example_query}
\end{example}

Example~\ref{ex:example_query} shows a graph traversal query
intended to find 20 users
who are within the 5-hop neighborhood of
user $123$, work at company `XYZ', and have tweeted with hashtag `\#ABC'.
When executing this query, we need to start a traversal for
every user entering the \textit{where} subquery.
This traversal can be canceled immediately if any tweet of the user
is found to have the desired hashtag, without needing to check all of that user's tweets.
Clearly, canceling this traversal should not affect the traversals for other users.
More importantly, it should not be blocked by other traversals,
e.g., by  someone who tweets a lot but not with hashtag `\#ABC'.
This implies that a GQS needs to support concurrent execution as well as
fine-grained control of subquery traversals (\textbf{O1-1}).

Eagerly checking the hashtags tweeted by one user
requires a DFS scheduling policy for the \textit{where} subquery.
The same scheduling policy holds within one loop iteration in
the \textit{repeat} subquery, as we would like to eagerly check if
a neighbor works at company `XYZ'. However, it would be preferable to use
BFS when scheduling across loop iterations of the \textit{repeat} subquery.
This is because we do not want to blindly explore all neighbors within 5 hops
if 20 candidates can be found in a much smaller neighborhood.
This implies that GQS systems need to support customized traversal policies in subqueries (\textbf{O1-2}).

\noindent\textbf{Performance Isolation and Load Balancing (O2).}
It is challenging to fulfill the stringent latency requirement in a
production environment.
Due to the intrinsic skewness in graph data,
graph traversal queries can vary dramatically in terms of the amount of computation.
Therefore, GQS should be
capable of enforcing performance isolation across queries to guarantee
the latency SLO.
In addition, as illustrated in Example~\ref{ex:example_query},
for better performance, the isolation granularity needs to be as small
as subquery-level traversals.
Existing graph databases~\cite{neo4j, neptune, janusgraph}
map concurrent queries to system threads and rely on the operating system for scheduling.
This mechanism does not expose the internal query complexity
and thus cannot support subquery-level isolation.
The skewness in graph data can lead to dynamic workload skewness at run time,
requiring GQS to provide dynamic load balancing.


Table~\ref{table:systems} compares existing graph query engines on their supports for
\textbf{O1} and \textbf{O2}.
Many graph query engines~\cite{neo4jplan, Mellbin, chen2019grasper, gaia, wukong,tigergraph}
use the dataflow model to execute graph traversal queries in languages
such as Cypher~\cite{cypher}, Gremlin~\cite{tinkerpop} and GSQL~\cite{tigergraph}.
Existing dataflow models either a) support only static topologies
(e.g., Timely~\cite{naiad} and GAIA~\cite{gaia}) or b) can only
dynamically spawn tasks at coarse granularity (e.g., CIEL~\cite{murray2011ciel}).
Specifically, Timely and GAIA attach a metadata
(\textit{timestamp} in Timely and \textit{context} in GAIA) in
every message to identify which subquery traversal it belongs to,
but then these systems process messages belonging to different traversals of a
subquery in the same static execution pipeline determined at compile time.
Without physically isolating the execution of traversals inside a subquery,
these systems cannot efficiently support both \textbf{O1} and \textbf{O2}.
Achieving these goals requires flexibly controlling and scheduling
the fine-grained subquery traversals spawned dynamically during query execution.
We expand more in Section~\ref{sec:motivation} to discuss the limitations of
existing dataflow models, and evaluate this issue with experiments in Section~\ref{sec:evaluation}.

In this paper, we propose a novel \textit{scoped dataflow} to model graph traversal queries,
which supports fine-grained control and scheduling during query execution.
The scoped dataflow model introduces a key concept called \textit{scope}.
A scope marks a subgraph in the dataflow, which corresponds to a subquery.
The scope can be dynamically replicated at runtime into physically isolated
scope instances which
correspond to independent traversals of the subquery.
Scope instances can be concurrently executed and independently controlled.
This way, traversals of a subquery can time-share the CPU and be independently canceled
without blocking or affecting each other.
Furthermore, a scope allows users to customize the scheduling policy
between and inside scope instances,
supporting diverse scheduling policies for different parts of a query.

We build the \textit{\gqs{}} engine for a multi-tenant GQS,
based on an efficient distributed implementation of scoped dataflow.
\gqs{} parallelizes a scoped dataflow into a physical plan of operators,
and cooperatively schedules these operators on executors pinned on physical cores.
On each executor, \gqs{} dynamically creates and terminates operators
to instantiate and cancel scope instances.
\gqs{} manages operators hierarchically as an operator tree:
operators are scheduled recursively by their parent scope operators.
This hierarchy allows customized scheduling
within each scope, and provides performance isolation
both across queries and within a single query.
\gqs{} partitions the graph into fine-grained \textit{tablets}
and distributes tablets across executors.
To handle workload skewness, \gqs{} dynamically
migrates tablets along with the operators accessing
them between executors for load balancing.
In summary, the contributions of this paper include:

\begin{enumerate}[nosep,leftmargin=*]
  \item We propose the scoped dataflow model, which introduces a novel scope construct to
  a dataflow. The scope explicitly exposes the concurrent execution and control of subgraphs
  in a dataflow to the finest granularity (Section~\ref{sec:scope}).

  \item We build \gqs{}, an engine for GQS based on a distributed
  implementation of the scoped dataflow model. \gqs{} can leverage the many-core parallelism
  in a modern server, and can easily scale out to a distributed cluster (Section~\ref{sec:haf}).

  \item We conduct extensive evaluations of \gqs{} on popular benchmarks.
  The results show that \gqs{} has 1-3 orders of magnitude performance improvement
  over the state-of-the-art graph query engines and provides performance
  isolation and load balancing.
\end{enumerate}

\begin{figure*}[tbh!]
\centering
\begin{tabular}{cccc}
\subfigure[]{
  \includegraphics[height=0.18\textwidth]{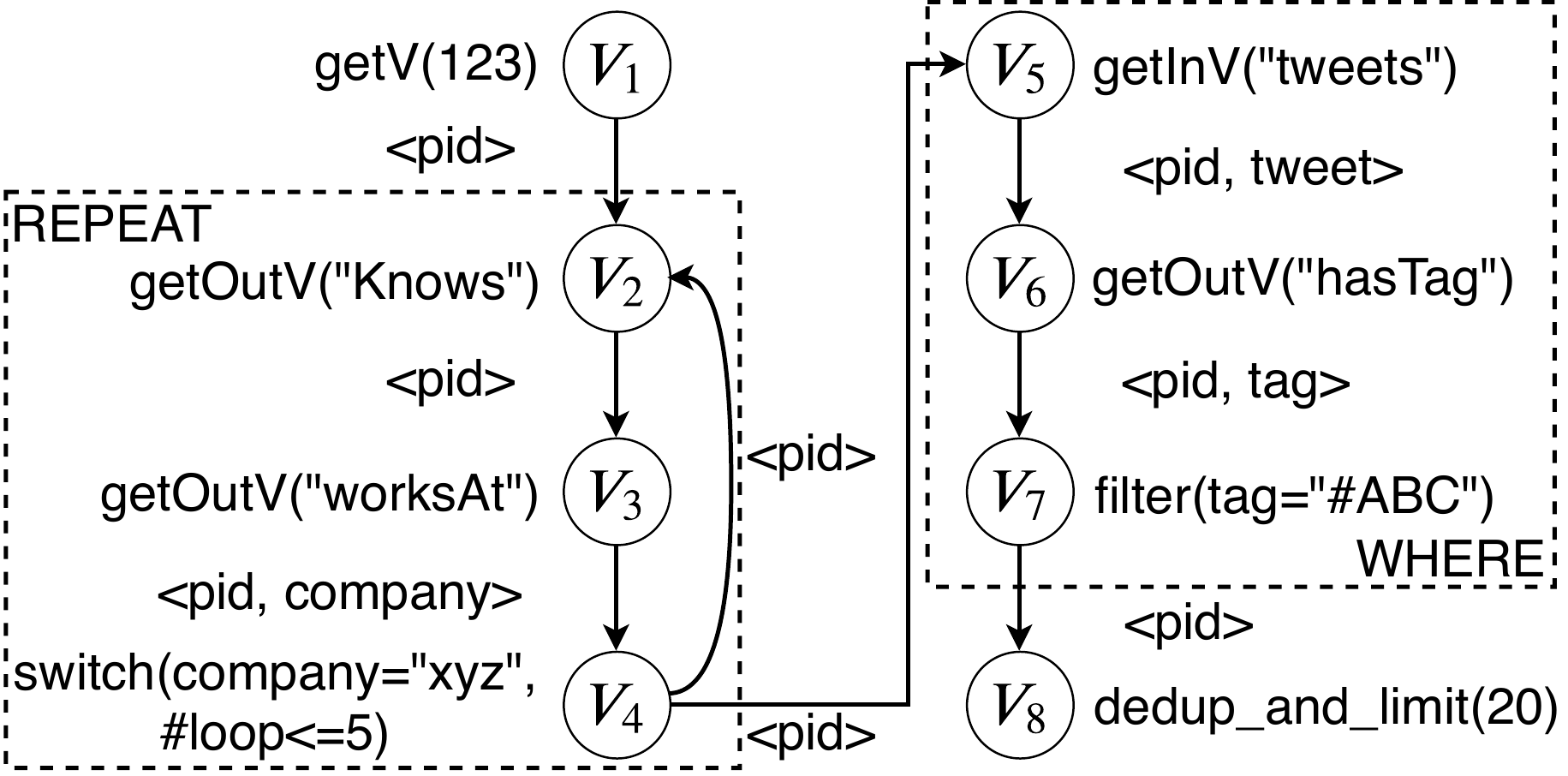}
  \label{fig:example-dataflow}
}
\hspace{-10pt}
&
\subfigure[]{
  \includegraphics[height=0.19\textwidth]{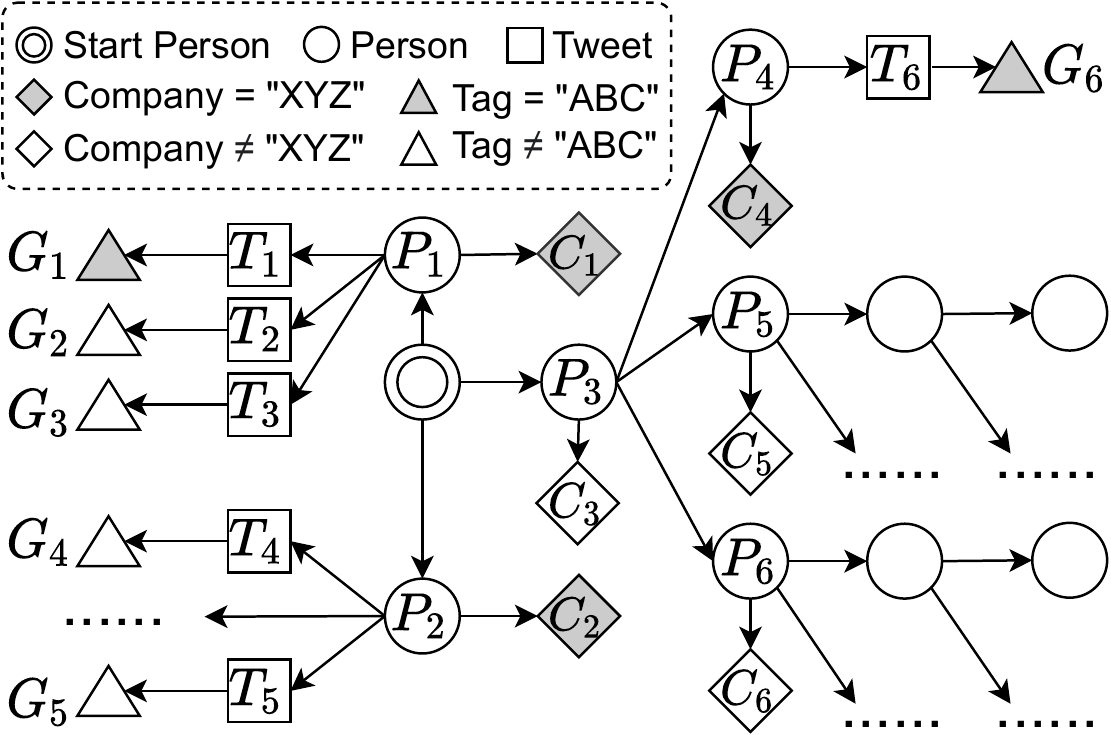}
  \label{fig:example-graph}
}
\hspace{-10pt}
&
\subfigure[]{
  \includegraphics[height=0.20\textwidth]{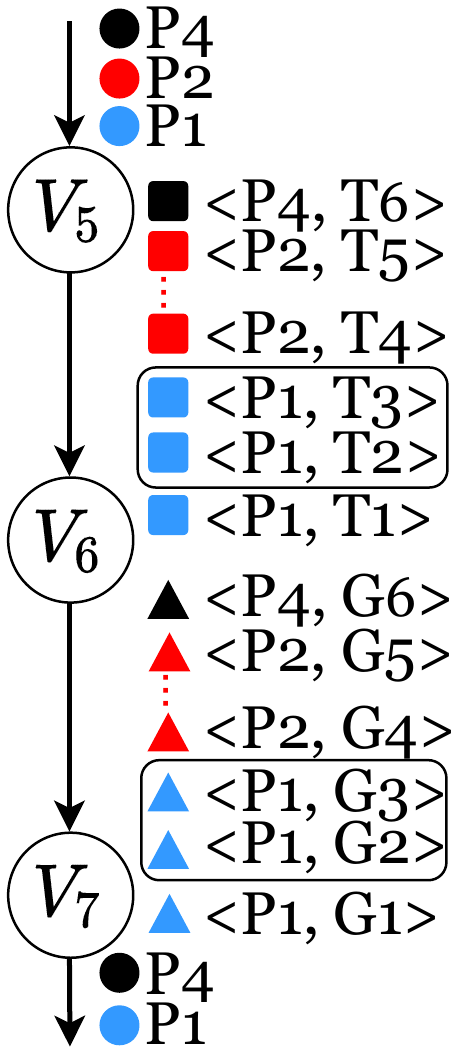}
  \label{fig:problem_1}
}
&
\subfigure[]{
  \includegraphics[height=0.20\textwidth]{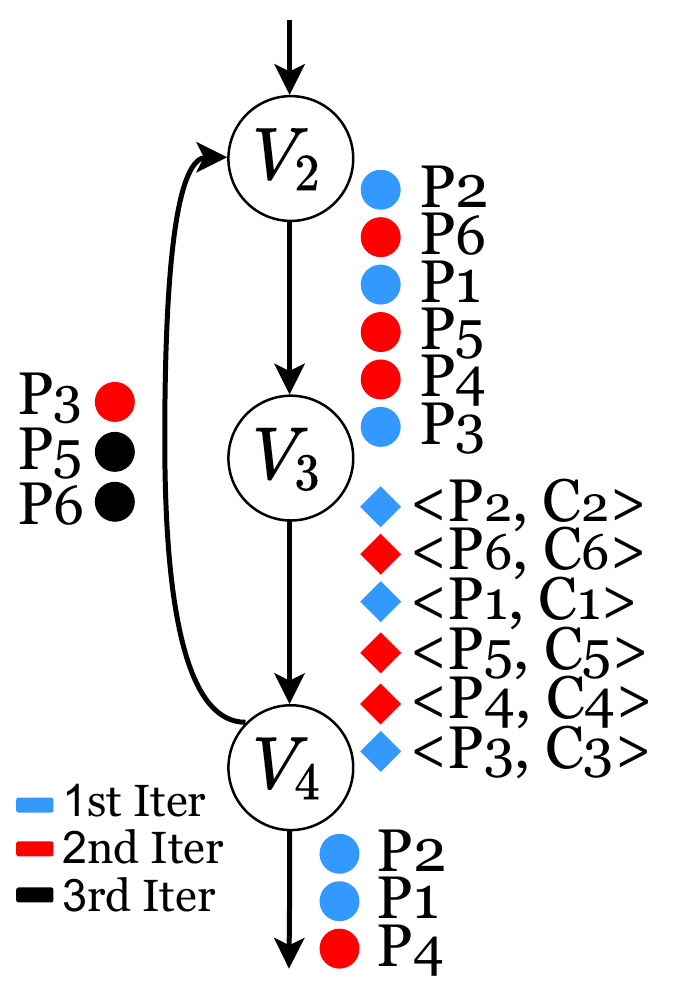}
  \label{fig:problem_2}
}
\end{tabular}
\vspace{-15pt}
\caption{(a) The logical dataflow for Example~\ref{ex:example_query}.
(b) An example data graph.
(c)(d) Execution pipelines of the \textit{where} and \textit{repeat} subquery
of the topo-static dataflow model in Figure~\ref{fig:example-dataflow}, respectively.
The execution pipelines depict snippets
of all messages generated during query execution at the operators.}
\vspace{-10pt}
\end{figure*}

\section{Limitations of Existing Dataflow Models}
\label{sec:motivation}

In the dataflow model, a graph traversal query is represented
as a directed graph of operators as vertices, where each
operator sends and receives messages along directed edges.
Figure~\ref{fig:example-dataflow} presents a typical logical
dataflow for Example~\ref{ex:example_query}, where
the \textit{repeat} subquery is translated into vertices $v_2$, $v_3$ and $v_4$,
the \textit{where} subquery is translated into vertices
$v_5$, $v_6$ and $v_7$\footnote{For simplicity, we omit projection operators in the dataflow,
and annotate the schema of the message along each edge.}.
Other dataflow models such as Timely~\cite{naiad} and GAIA~\cite{gaia} share the same main structure
demonstrated in Figure~\ref{fig:example-dataflow}. A common characteristic of these dataflow
models is that their topologies are fixed at compile time.
We refer to such dataflow models as \textit{topo-static} dataflows in the
following discussions. Next, we use Example~\ref{ex:example_query} on the data graph
in Figure~\ref{fig:example-graph} to illustrate the limitations of using topo-static dataflows in a GQS.

\noindent\textbf{Control on Concurrent Traversals (O1-1).}
A graph query usually has subqueries that can launch
many independent traversals starting
from different vertices in the graph.
E.g., each sub-tree rooted at $P_1$, $P_2$, $P_4$ and so on in
Figure~\ref{fig:example-graph} are independent traversals of the \textit{where} subquery.
Figure~\ref{fig:problem_1} demonstrates an example execution pipeline of the
\textit{where} subquery in the topo-static dataflow model in
Figure~\ref{fig:example-dataflow}, where messages of different traversals
(marked in different colors) are mixed in the dataflow and
sequentially processed. Although the traversal of $P_1$ can be
immediately terminated when $V_7$ detects $G_1$,
other messages in the traversal of $P_1$ (messages in black rectangles)
cannot be trivially canceled:
To cancel a specific traversal, one has to annotate each message with
extra metadata about which traversal it belongs to, and filter messages
by their metadata at operator $V_5$-$V_7$.
GAIA cancels subquery traversals in this way.
Alternatively, simply isolating the traversals using different threads will not work.
A large number of concurrent traversals can lead to an exploded number of threads,
where the context switch overhead will set a hard bottleneck on
the system performance. This impact becomes even worse in a multi-tenant environment.

\noindent\textbf{Diverse Scheduling Policies (O1-2).}
Inside a graph query, different subqueries often have very diverse
scheduling preferences. Consider the \textit{repeat} subquery.
Instead of blindly exploring all the 5-hop neighbors, we prefer to gradually expand
the exploration radius, as closer neighbors (e.g., $P_1$ and $P_2$ in the first hop)
are more likely to work at the same company with the start person.
This corresponds to completing the traversals of
earlier iterations first, i.e., in a BFS manner.
Meanwhile, inside an iteration we prefer to
finish checking if a neighbor is a match before the next, i.e., scheduling operators
$V_2$-$V_4$ in a DFS manner.
However, as depicted in Figure~\ref{fig:problem_2},
in the topo-static dataflow model,
messages of different iterations are mixed and may be disordered due to parallel
execution, e.g., $P_4$ and $P_5$ are processed
before $P_1$ in $V_3$. To enforce inter-iteration BFS,
one has to annotate each message with its belonging iteration,
and sort every incoming message according to this metadata in $V_2$-$V_4$,
incurring much overhead. As far as we know, no existing graph query engine
allows configuring subquery-level scheduling policies.

\noindent\textbf{Performance Isolation (O2).}
In a GQS, the scales of traversals
could vary drastically between queries. Even with the
same query, different inputs could lead to traversals of very different scales.
E.g., on the LDBC benchmark~\cite{ldbc}, we observed up to three orders of
magnitude difference in query latency for the same query with different starting persons.
In addition, performance isolation is necessary for concurrent traversals
inside subqueries to prevent a traversal with heavy computation
from indefinitely blocking other traversals.
E.g., in Figure~\ref{fig:problem_1} the traversal of $P_2$ (messages in red),
who tweeted a lot without hashtag \textit{\#ABC}, blocks the traversal of $P_4$ (messages in black),
even if $P_4$ can pass the predicate. This requires the execution framework to provide performance isolation
in various granularities, from the level of inter-user in a multi-tenant
environment, to the level of inter-traversal within a subquery.

\section{Scoped Dataflow}
\label{sec:scope}

\textit{Scoped dataflow} is a new computational model that extends the existing dataflow model,
with the ability to explicitly expose concurrent execution and control of subgraphs in a dataflow to the finest granularity.
In this section, we define the structure of scoped dataflow, introduce the
programming model, and demonstrate scoped dataflow can tackle
the problems in Sec~\ref{sec:motivation}.

\subsection{Computation Model}
\label{sec:model}

\begin{figure*}[tbh!]
\centering
\begin{tabular}{ccccc}

\subfigure[]{
  \includegraphics[height=0.17\textwidth]{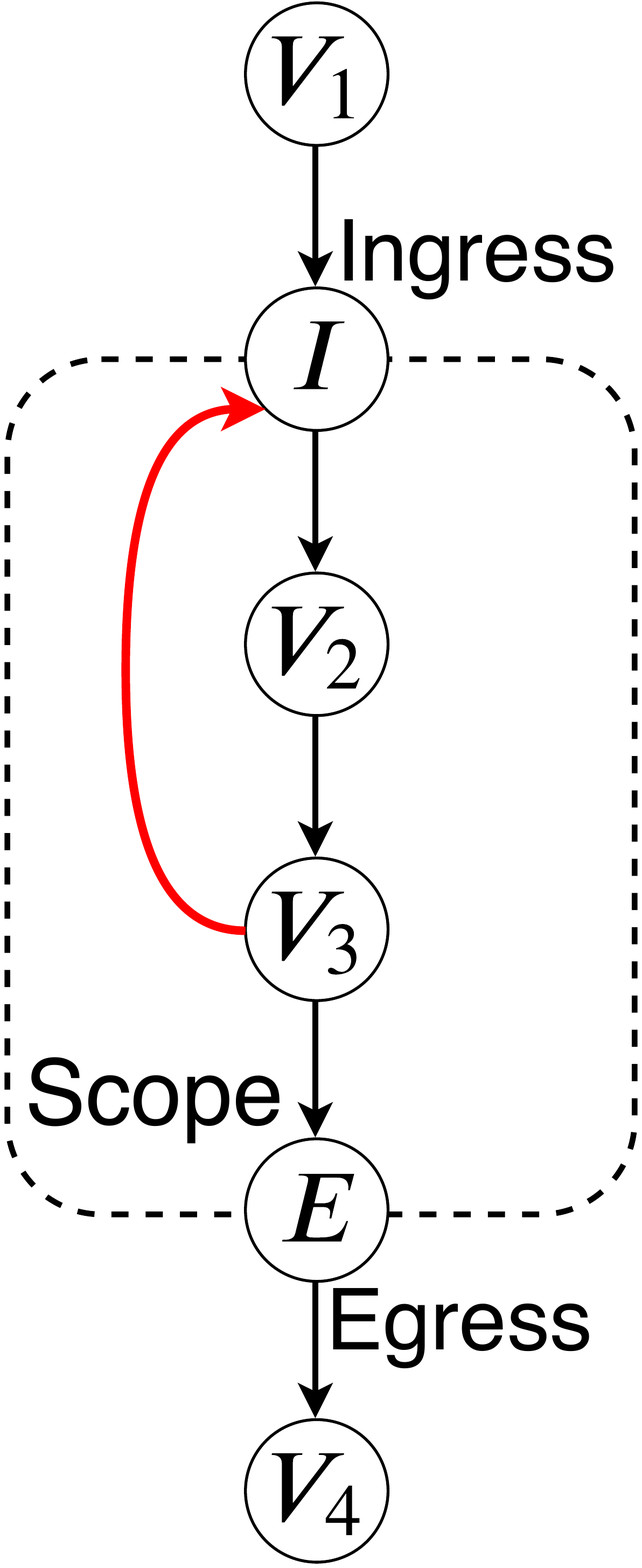}
  \label{fig:example-scoped}
}
&
\subfigure[]{
  \includegraphics[height=0.17\textwidth]{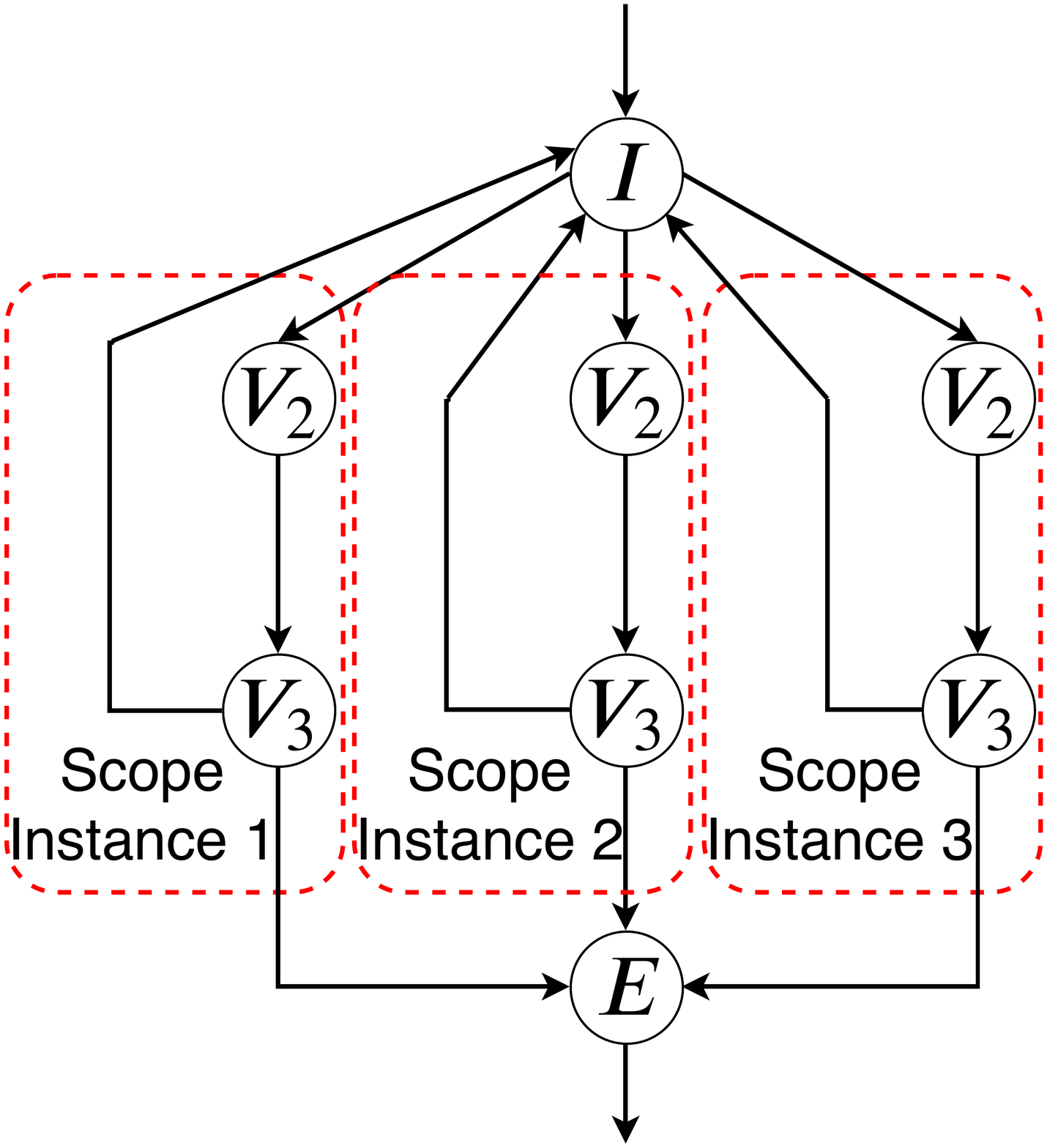}
  \label{fig:instantiated-scoped}
}
&
\subfigure[]{
  \includegraphics[height=0.17\textwidth]{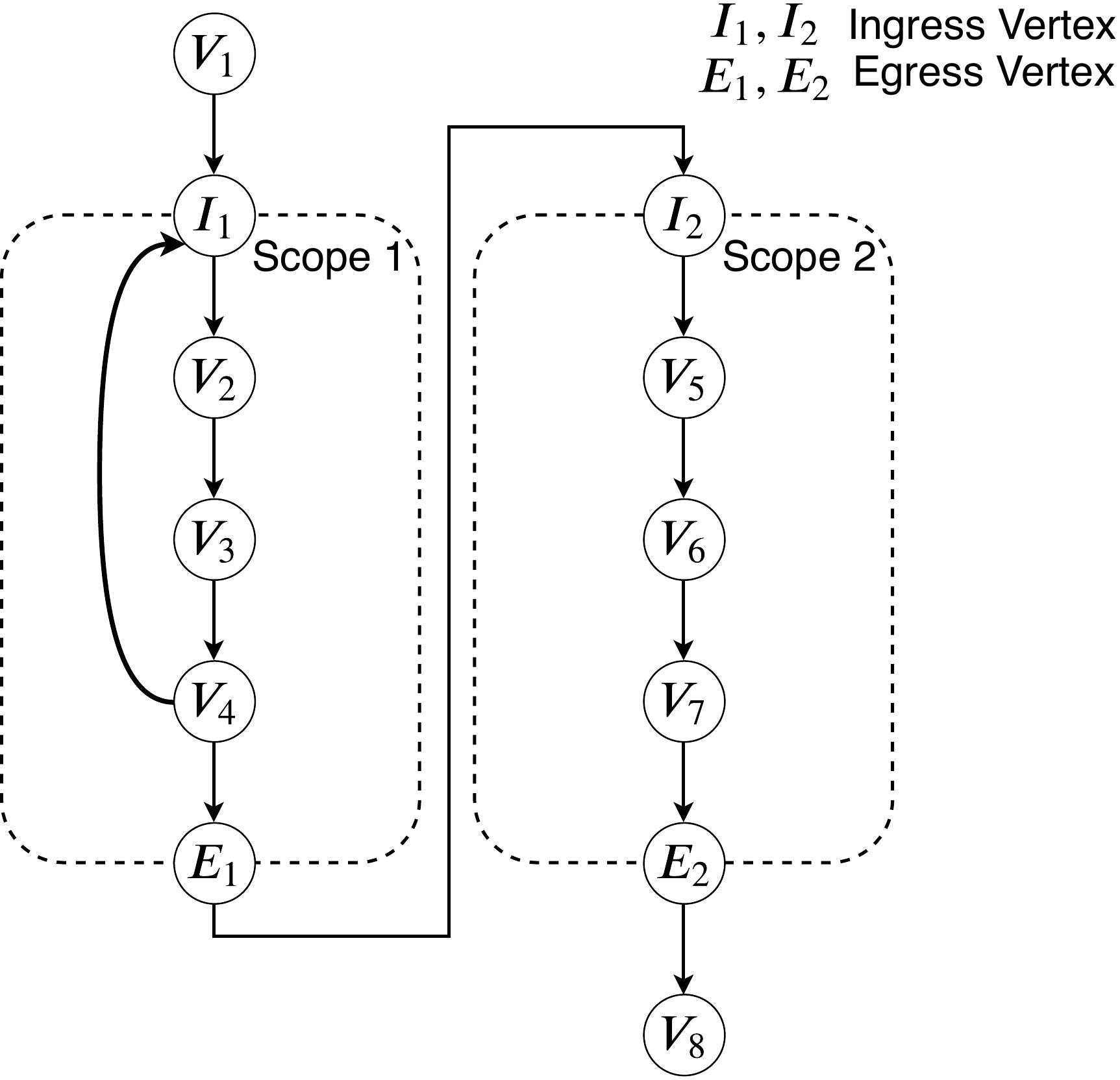}
  \label{fig:example-solution}
}
&
\subfigure[]{
  \includegraphics[height=0.17\textwidth]{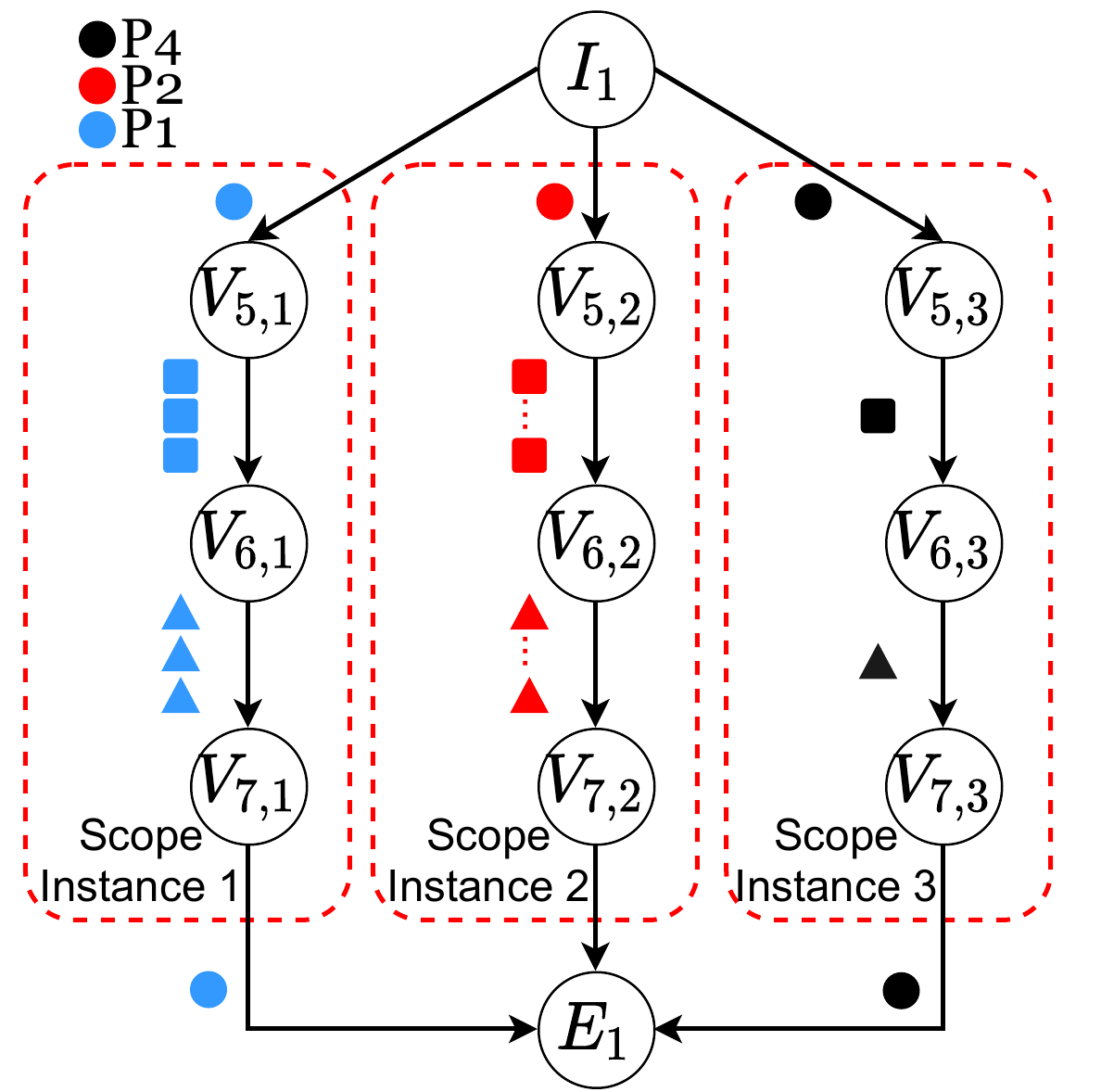}
  \label{fig:solution_branch}
}
&
\subfigure[]{
  \includegraphics[height=0.17\textwidth]{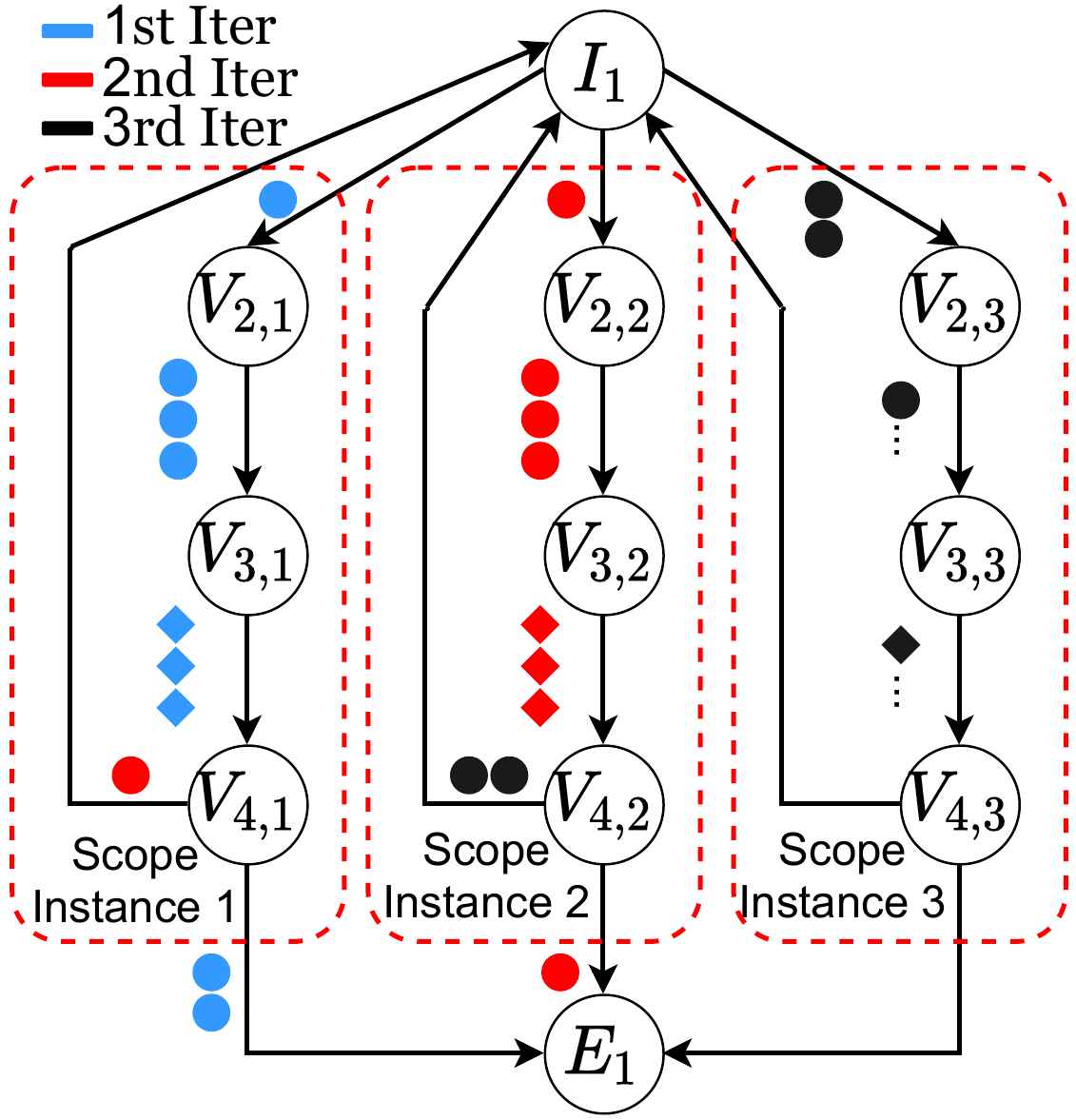}
  \label{fig:solution_loop}
}
\end{tabular}
\vspace{-15pt}
\caption{
(a) An example loop scope,
and (b) an example of its instantiation.
(c) The scoped dataflow for Example~\ref{ex:example_query},
and an instantiation of its (d) branch scope and (e) loop scope.}
\vspace{-10pt}
\end{figure*}

Similar to the traditional dataflow model,
a scoped dataflow is also based on a directed graph $G(V, E)$.
Vertices in $V$ send and receive messages along directed edges in $E$.
The scoped dataflow introduces a new construct named
\textit{scope}. Formally, a scope is a sub-structure of the scoped
dataflow $G(V, E)$, and has two system-provided vertices:
an \textit{ingress vertex} and an \textit{egress vertex}.
All the input messages entering a scope pass through its ingress vertex, and all
the edges leaving a scope pass through its egress vertex.
Inside a scope $S$, vertices which are neither the ingress nor egress of $S$
are referred to as \textit{internal vertices} of $S$.
An internal vertex of $S$ can belong to an inner scope of $S$.
If $G(V, E)$ is cyclic, every cycle in $G(V, E)$ must be contained entirely within a scope $S$,
and the backward edge must be from a vertex $v$ in $S$, to the ingress vertex of $S$.
Since the edges leaving a scope must pass through its egress vertex,
$v$ cannot be in any inner scope of $S$.
We categorize scopes into two types:
a scope without backward edges is called a \textit{branch scope}, and a scope with backward edges
is called a \textit{loop scope}.
Figure~\ref{fig:example-scoped} shows an example loop scope.
Scopes can be well-nested. The nesting level of a scope is
called its \textit{depth}. The depth of a top-level scope is $1$.

A scope marks a region inside a dataflow:
the dataflow subgraph inside a scope can be dynamically replicated at runtime
to create new subgraph instances, isolating the processing of
different input data entering the scope. The newly instantiated dataflow subgraph of a scope
is called a \textit{scope instance}. The states of vertices in different scope
instances are independent. In a scope $S$, scope instances are instantiated
as follows:
\begin{itemize}
  \item Every inner vertex of $S$ and every edge connecting these
        inner vertices are copied in the new instance. Note that,
        the ingress/egress vertices of any inner scopes contained in $S$
        are also counted as $S$'s inner vertices. The state of each
        copied stateful vertex is initialized as the default value.
  \item Every edge connecting $S$'s ingress/egress vertex and an inner vertex,
        including backward edges, is copied
        by replacing the inner vertex with its corresponding copy in the new instance.

\end{itemize}
\noindent
Figure~\ref{fig:instantiated-scoped} shows an example of instantiated scoped dataflow
with three scope instances for Figure~\ref{fig:example-scoped}.

For a scoped dataflow $G(V,E)$, we denote its instantiated scoped dataflow as
$\tilde{G}(\tilde{V}, \tilde{E})$. In $\tilde{G}$, each vertex (resp. edge) in a scope instance
can be uniquely identified by the corresponding $v$ (resp. $e$) in $G$,
and a \textit{scope tag} $t$ of the containing scope instance.
The \textit{scope tag} is in the following format:

$$ScopeTag: \langle s_1, \cdots, s_d \rangle \in \mathbb{N}^d$$
where an element $s_k$ denotes the $s_k$-th scope instance in a scope
of depth $k$. The vertex (resp. edge) in $\tilde{G}$ identified by $v$ (resp. $e$) and the scope
tag $t$ is denoted by $\tilde{v}_t$ (resp. $\tilde{e}_t$).
Vertices and edges not in any scope have an empty scope tag $\langle \rangle$.

\noindent\textbf{Programming Model.}
In a scoped dataflow, every message bears the scope tag of the edge it passes through.
Every vertex implements the following APIs:

\begin{align*}
&v.\text{\textit{ReceiveMessage}}(msg : Message, e : Edge, t : ScopeTag) \\
&v.\text{\textit{OnCompletion}}(t : ScopeTag)
\end{align*}
A vertex may invoke two system-provided methods in the context of the above callbacks:

\begin{align*}
&this.\text{\textit{SendMessage}}(msg : Message, e : Edge, t : ScopeTag) \\
&this.\text{\textit{NotifyCompletion}}(t : ScopeTag)
\end{align*}
Users write their logics according to the scoped dataflow $G$,
while at runtime the logics are executed by $\tilde{G}$.
For instance, $v.$\textit{ReceiveMessage(msg, $e$, $t$)}
defines the logics how $\tilde{v}_t$ processes a message received
along $\tilde{e}_t$.
$v.\text{\textit{OnCompletion}}(t)$ defines the logics executed
when vertex $\tilde{v}_t$ has no more input, e.g., an \textit{aggregate}
operator emits the final aggregation results when it sees all the inputs.
$\text{\textit{SendMessage}}(msg, e, t)$
sends a message along with edge $\tilde{e}_t$.
$\text{\textit{NotifyCompletion}}(t)$ can be called if an operator would
like to terminate processing proactively, e.g., a \textit{limit} operator terminates
once it generates enough outputs.

\noindent\textbf{Scope Instantiation.}
In a scope, the ingress vertex instantiates scope instances
and routes the messages entering this scope to different scope instances. The egress
vertex manages the termination of scope instances inside a scope.
Scope instances can be terminated independently of each other.
The ingress and egress vertices only act on the scope tags of messages passing through.
In specific, for each message $msg$ passing through:

\begin{itemize}
  \item The ingress vertex routes $msg$ to a destination scope instance $SI$ and
  sets the scope tag of $msg$ to that of $SI$. If $SI$ does
  not exist, the ingress vertex instantiates it.

  \item The egress vertex removes the last element in the scope tag of $msg$.
  When \textit{OnCompletion($t$)} is called in the egress, it terminates the
  scope instance with scope tag $t$.
\end{itemize}


Branch and loop scopes have different behaviors on how messages
are mapped to scope instances. In a branch scope, every input message
triggers the instantiation of a new scope instance. Whereas in a loop scope,
messages from edges entering the scope with scope tag $\langle s_1, \cdots, s_d \rangle$
are routed to the scope instance with scope tag $\langle s_1, \cdots, s_d, 1 \rangle$;
messages from backward edges with scope tag $\langle s_1, \cdots, s_d, s_{d+1} \rangle$
are routed to scope instance with scope tag $\langle s_1, \cdots, s_d, s_{d+1} + 1 \rangle$.
A configurable threshold, \textit{Max\_SI}, can be set to constrain the
maximal number of concurrent scope instances in a scope.

\noindent\textbf{Scope Scheduling.}
Scoped dataflow supports customizing scheduling policies for scopes.
The scheduling policy of a scope can be decoupled into two parts: \textit{inter-scope-instances}
(inter-SI) policy and \textit{intra-scope-instance} (intra-SI) policy.
The inter-SI policy specifies the scheduling priorities of scope instances
inside a scope. The intra-SI policy specifies the scheduling priorities
of inner vertices (an inner scope as a whole is treated as a virtual inner vertex)
inside a scope instance. Users can customize the scheduling policies with the below comparators:

\begin{align*}
&bool ~ \text{\textit{InterSI\_Comparator}}(t_1 : ScopeTag, t_2 : ScopeTag) \\
&bool ~ \text{\textit{IntraSI\_Comparator}}(v_1 : VertexID, v_2 : VertexID)
\end{align*}
\vspace{-10pt}

The inter-SI comparator decides the priorities of scope instances.
The intra-SI comparator decides the priorities of inner vertices in the same scope
in $G(V,E)$.

For any two vertices $\tilde{u_{t_1}}$ and $\tilde{v_{t_2}}$ in $\tilde{G}$,
their scheduling orders are decided iteratively according to the following rules.
Without loss of generality, we denote the depths of $t_1$ and $t_2$
as $d_1$ and $d_2$, and assume $d_1 \leq d_2$. We use $anc_d(\tilde{v_t})$ to
denote the ancestor scope instance of $\tilde{v_t}$ at depth $d$.

\begin{itemize}
  \item We start comparing the ancestor scope instances of $\tilde{u_{t_1}}$
  and $\tilde{v_{t_2}}$ from depth $1$ to depth $d_1$.

  \item At depth $d$, if $anc_d(\tilde{u_{t_1}})$ and $anc_d(\tilde{v_{t_2}})$
  are the same, we proceed to depth $d+1$.

  \item At depth $d$, if $anc_d(\tilde{u_{t_1}})$ and $anc_d(\tilde{v_{t_2}})$
  are different scope instances of the same scope $S$, the priority is determined
  by calling the inter-SI comparator of $S$ on the scope tags of
  $anc_d(\tilde{u_{t_1}})$ and $anc_d(\tilde{v_{t_2}})$.

  \item At depth $d$, if $anc_d(\tilde{u_{t_1}})$ and $anc_d(\tilde{v_{t_2}})$
  belong to different scopes in a common parent scope $S$,
  the priority is determined by calling the intra-SI
  comparator of $S$ on the scopes of $anc_d(\tilde{u_{t_1}})$ and $anc_d(\tilde{v_{t_2}})$.
\end{itemize}

\subsection{Progress Tracking}
\vspace{-0.3em}
\label{sec:progress}
To correctly invoke $\text{\textit{OnCompletion}}(t)$ for vertices,
a scoped dataflow needs to track the processing progress of its vertices,
i.e., when a vertex is guaranteed to have received all its inputs.
The scoped dataflow model adopts an EOS-based progress tracking mechanism similar to the
Chandy-Lamport algorithm~\cite{chandy}.  After ingesting all the external
inputs, the runtime automatically inserts an EOS message to the dataflow.
EOS messages are propagated through the dataflow graph to facilitate
the progress tracking. Once a vertex receives EOS messages from all
its incoming edges, it calls $\text{\textit{OnCompletion}}(t)$, and then
emits an EOS message in all its outgoing edges.

However, as a scoped dataflow can dynamically instantiate scope instances
and may contain cycles, we extend the aforementioned EOS-based progress tracking
mechanism in order to support the scoped dataflow.


\noindent\textbf{Hierarchical Progress Tracking.}
We track progress hierarchically in a scoped dataflow, i.e.,
progress tracking inside and outside a scope are conducted separately.
Progress tracking outside a scope $S$ simply treats $S$ as a virtual vertex, denoted $v_S$.
The runtime conducts progress tracking on the dataflow subgraph inside a scope $S$
to decide the completion of $v_S$, which completes
when the ingress vertex of $S$ receives EOS from all its input edges,
and all the scope instances inside $S$ have completed.
Once $v_S$ reaches completion, the egress vertex of $S$
emits EOS along all its outgoing edges.

\noindent\textbf{Tracking inside a Scope.}
Next, we explain how the progress tracking is done inside the branch and loop scope, respectively.

In a branch scope, the runtime propagates EOS in each scope instance
to track their progress independently.
An ingress vertex reaches completion when it has received EOS from all its incoming edges.
After completion, the ingress vertex sends the largest scope instance ID
it has spawned to the egress vertex of the same scope.
The egress decides the number of scope instances
according to this ID.
When the egress vertex has tracked that all the scope instances in this scope have
completed, it reaches completion and emits EOS to the outgoing edges.

Progress tracking of loop scopes is extended from that of the branch scopes.
To decide when the loop iteration reaches completion,
the ingress vertex tracks the scope tags of data and EOS messages
received from the backward edges. If an ingress vertex only receives
EOS but no data messages for a specific scope instance, the ingress vertex
can infer that this scope instance is the last loop iteration.
This way, the ingress vertex can infer the number of spawned scope instances.
With this information, the egress vertex tracks the progress
of scope instances in the same way as the branch scope.
This process is guaranteed to be able to stop due to the simple fact
that if you remove the ingress vertex inside a loop scope (note that the ingress
vertex only forwards messages) and directly connect the edges according
to the forwarding behavior of the ingress vertex,
the instantiated scope dataflow is a DAG without cycles.

\subsection{Scoped Dataflow in Action}
In this section, we discuss the applicability of scopes,
followed by an example explaining how the scoped dataflow model can be used to solve
the challenges discussed in Section~\ref{sec:motivation}.

\noindent\textbf{Applicability of Scopes.}
The scoped dataflow model is designed to facilitate fine-grained control on
subquery traversals and enforce scope-level customization of scheduling policies.
In general, scopes can benefit graph queries with:
\begin{itemize}
\item \textit{Where} subqueries which can be early canceled.
As the instantiation of scope instance brings overhead
(see \textbf{E2} in Section~\ref{subsec:expr-scope}).
For \textit{where} subqueries which cannot be early canceled,
scopes should be turned off.

\item \textit{Loop} subqueries that can find matches more quickly
following certain exploration strategy (e.g., BFS or DFS).
\end{itemize}
Note that, the benefit of scope is query/data-dependent (see \textbf{E2} in
Section~\ref{subsec:expr-scope}). The best plan should be determined by the query compiler,
which is beyond the scope of this paper.

\begin{example}[Implementation of vertex $V_7$ in Figure~\ref{fig:solution_branch}]
\noindent

\begin{lstlisting}[style=C++Style, upquote=false]
class V7Filter:Vertex {
  void ReceiveMessage(msg:Message, e:Edge, t:ScopeTag) {
    if (msg.GetTag() == "#ABC") {
      SendMessage(msg.getPersonId(), out_e, t);
      NotifyCompletion(t);}
  }
  void OnCompletion(t:Tag) { }
}
\end{lstlisting}
\label{ex:cancel-code}

\end{example}

\begin{example}[Implementations of the inter-SI BFS and intra-SI DFS policies]
\noindent

\begin{lstlisting}[style=C++Style, upquote=false]
bool InterSI_BFS:InterSI_Comparator(t_1:ScopeTag, t_2:ScopeTag) {
  return LexicalOrderCompare(t_1, t_2);
}
bool IntraSI_DFS:IntraSI_Comparator(v_1:VertexID, v_2:VertexID) {
  return v_1 < v_2;
}
\end{lstlisting}
\label{ex:policy-impl}

\end{example}

\noindent\textbf{Examples of Scopes.}
Figure~\ref{fig:example-solution} shows the scoped dataflow for
Example~\ref{ex:example_query}. Figure~\ref{fig:solution_branch} zooms in
the \textit{where} subquery of Figure~\ref{fig:example-solution}, and
demonstrates instantiations of scope $2$. Traversals triggered by
different users entering the \textit{where} subquery are mapped to
different scope instances, which can be executed concurrently and controlled
independently. This way, a user (e.g., the blue one) who posts many tweets without
the specified tag will not block the exploration of other users (e.g., the red
and black ones) entering the \textit{where} subquery.

Example~\ref{ex:cancel-code} shows the implementation of
the \textit{filter} vertex $v_7$ in Figure~\ref{fig:solution_branch},
which enables early cancellation:
On receiving a message $msg$, if the tag in $msg$
is a match, the vertex notifies the completion of itself by calling
$\text{\textit{NotifyCompletion}}(t)$ to trigger its cancellation.
As a scope instance can be terminated independently,
when a match (the red message marked in black box entering $v_{7,2}$)
is found, the corresponding scope instance ($v_{5,2},
v_{6,2}, v_{7,2}$) can be canceled without impacting the other scope instances.

The scheduling policies of scopes can be flexibly configured to
fulfill the diverse scheduling preferences of different parts
in a graph query. Figure~\ref{fig:solution_loop} shows that iterations
of the \textit{repeat} subquery in Figure~\ref{fig:example-solution} are mapped into different
scope instances of the loop scope. We configure a \textbf{BFS} inter-SI scheduling policy
such that the blue scope instance (the first iteration) is executed first, then the red one
(the second iteration), and the black one (the third iteration) as the last.
Meanwhile, by enforcing a \textbf{DFS} intra-SI policy,
vertices inside the blue scope instance are scheduled in the order of
$\tilde{v}_{4,1}$, $\tilde{v}_{3,1}$ and $\tilde{v}_{2,1}$.
Implementations of the inter-SI BFS
and intra-SI DFS policy are presented in Example~\ref{ex:policy-impl}.

\section{Building \gqs{} on Scoped Dataflow}
\label{sec:haf}

We build \gqs{}, an engine for GQS based on a distributed implementation of
the scoped dataflow model. \gqs{} is designed to efficiently leverage the many-core
parallelism in a modern server, and agilely balance the workloads across cores.
\gqs{} can also easily scale out to a distributed cluster.

\begin{figure}[t]
\centering
\vspace{-10pt}
\begin{tabular}{c}

\subfigure[]{
  \includegraphics[height=0.18\textwidth]{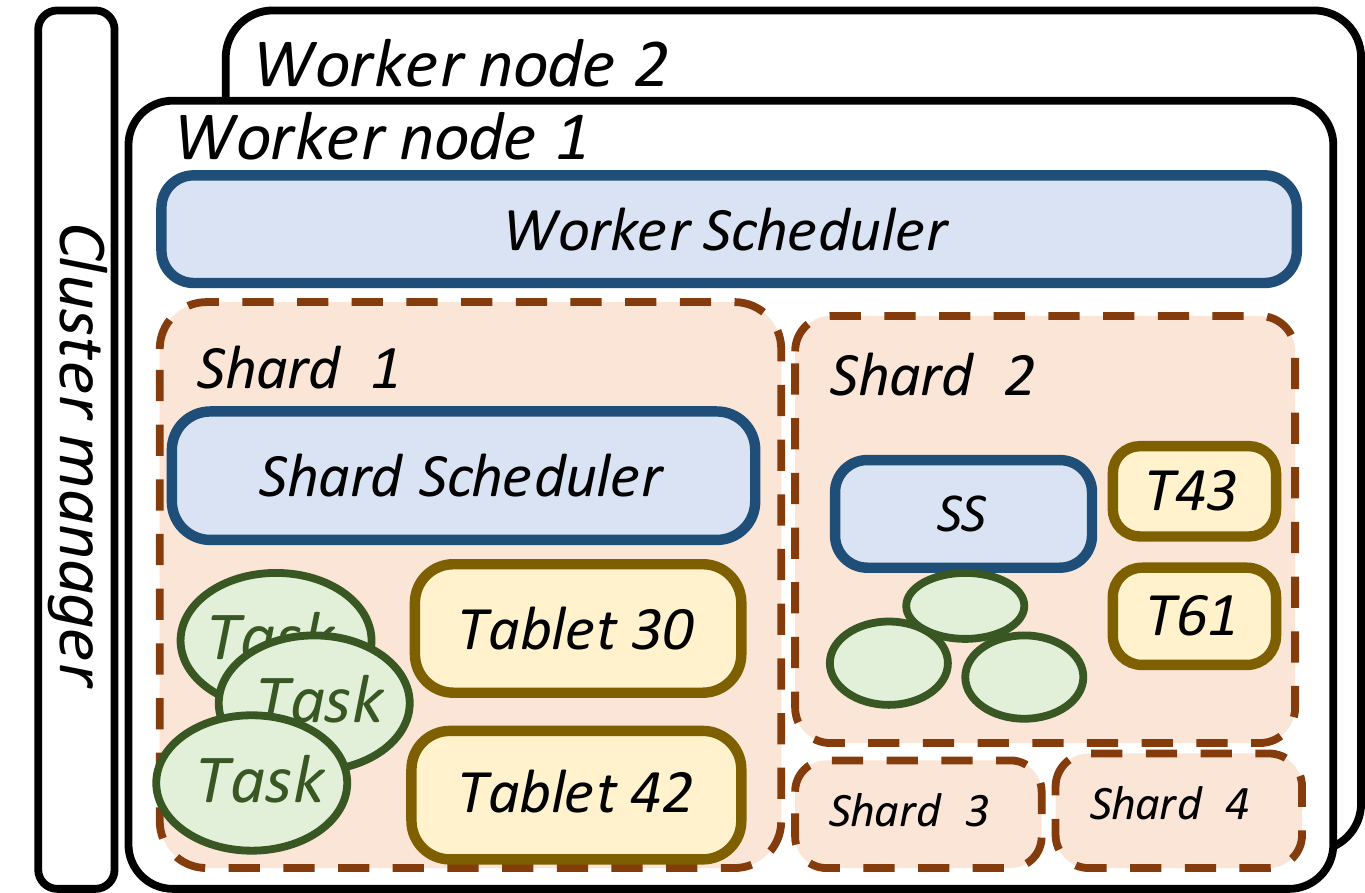}
  \label{fig:architecture}
}

\\

\subfigure[]{
  \includegraphics[height=0.15\textwidth]{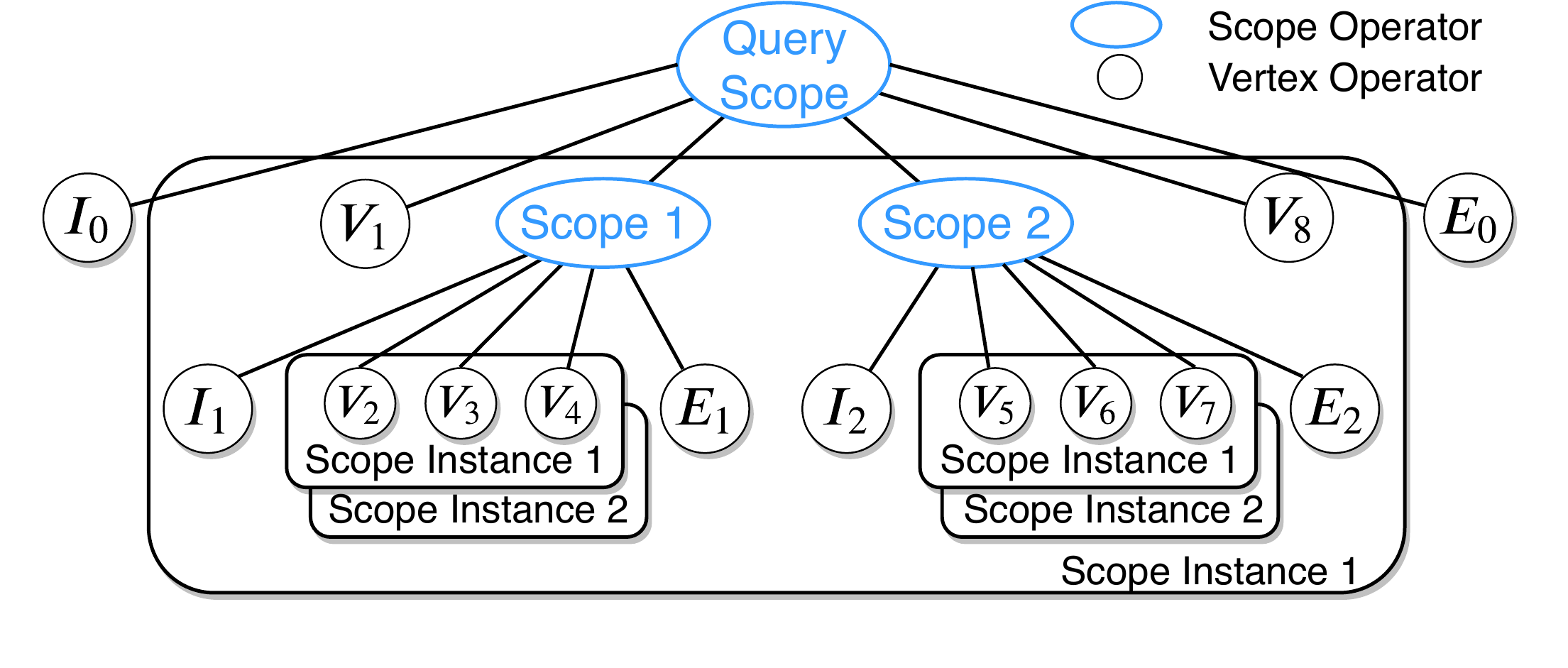}
  \label{fig:actor-ag}
}
\end{tabular}
\vspace{-15pt}
\caption{
(a) The system architecture of \gqs{}.
(b) An operator tree mapped from the dataflow in Figure~\ref{fig:example-solution} in an executor.
}
\end{figure}

The overall system architecture of \gqs{} is presented in Figure~\ref{fig:architecture}.
A \gqs{} cluster consists of a group of worker nodes, each of which manages
multiple executors. An executor exclusively runs in a system
thread pinned in a physical core, and is in charge of a partition of the graph data.
Vertices in a scoped dataflow are parallelized into operators, and are mapped to
executors. Executors communicate with each other through message queues inside the worker node and through
TCP connections across worker nodes.  In a worker node, executors schedule their
own operators and are scheduled by the worker scheduler. The worker scheduler
is responsible for balancing the workloads across executors in a worker node.

\subsection{Parallelizing A Scoped Dataflow}
\label{sec:parallelism}

\gqs{} parallelizes a dataflow $G$ into a physical plan of operators
in a data-parallel manner.
Every vertex in $G$ is parallelized into
a set of operators, each encapsulating the processing logic of
this vertex. Each edge in $G$ has a partitioning function controlling
the data exchange between the operators. In the physical plan, the
vertices in $G$ are replaced with the corresponding set of operators,
and the edges in $G$ are replaced with a set of edges connecting these operators.
At runtime, when we instantiate a scope in $G$ into scoped instances,
every scope instance inherits the physical plan of the scope.
That is, every scope instance has a complete copy
of the corresponding operators and edges of its scope.

\gqs{} partitions the graph into tablets distributed across executors.
A tablet contains an exclusive set of graph vertices and all their in/out edges,
along with their properties. We refer to vertices in $G$ that need to
access graph data as \textit{graph-accessing} vertices. To exploit data locality,
a graph-accessing vertex is parallelized into as many operators as the number
of tablets, and are mapped to the executors hosting the tablets.

This mechanism fits well with the NUMA architecture.
As graph-accessing operations are often highly random,
by collocating each \textit{graph-accessing} operator and the corresponding tablet in
one executor, we can utilize the low-latency intra-node random
memory access in NUMA servers~\cite{Zhang2015}. Messages passing across
executors are batched to utilize the high bandwidth of sequential memory access across
NUMA nodes.

\subsection{Executor Internals}
\label{sec:executor}

In \gqs{}, executors are single-threaded and each one is pinned in a physical core.
Executors schedule their operators cooperatively, allowing
\gqs{} to concurrently process a large number of operators without facing the bottleneck
caused by context switch.
Cooperative scheduling is based on an asynchronous task-based programming interface.
E.g., an operator blocked by an asynchronous I/O operation automatically
yields CPU, and the corresponding executor schedules another operator ready for
execution. This way, \gqs{} can overlap CPU computation with networking and I/O, and thus improve the resource utilization.

\noindent\textbf{Scope Operator.}
The key difference between the scoped dataflow and the traditional dataflow is
the introduction of the scope abstraction. To facilitate the scope-based scheduling
in \gqs{} (see Section~\ref{sec:model}), we deliberately introduce \textit{scope operators}
to manage the creation, termination and scheduling for all the operators of a scope.
In specific, on every executor containing operators of a scope $S$,
we create a scope operator managing all the local operators of $S$ in this executor.
The scope operator of $S$ is also managed by the scope operator of $S$'s parent scope (if existed).
Therefore, all the operators in an executor can be managed as a forest of operator trees.
In each tree, the leaf nodes are operators of vertices and the non-leaf
nodes are the scope operators.
The operators in an executor are scheduled
hierarchically: (1) The executor schedules the root scope operators of queries.
(2) When a scope operator of scope $S$ is scheduled, it further schedules its child (scope) operators following $S$'s inter-SI and intra-SI scheduling policies.

\gqs{} enforces the performance isolation guarantee
in each single executor. We take scope operator as the basic unit of resource allocation:
resources allocated to scopes operators at the same depth are isolated, and an operator
can only consume the resources allocated to its parent scope operator.
Once being scheduled, an operator is assigned a \textit{quota CPU time} by its parent,
which constrains the maximum amount of CPU time this operator can use at this round
of scheduling,
and yields immediately after using up its quota.
The vertex operator updates its quota after processing a message, and thus
will not occupy an executor for a long time.

By modeling different queries or tenants as the top-level scopes,
\gqs{} can naturally support performance isolation across queries or users.
Figure~\ref{fig:actor-ag} shows the operator tree
mapped from the scoped dataflow in Figure~\ref{fig:example-scoped}.

\subsection{Hierarchical Operator Management}

In this subsection, we introduce how operators are addressed, created
and terminated in \gqs{}.

\noindent\textbf{Operator Addressing.}
In \gqs{}, each operator has a unique address, encoding the path from
the executor to this operator in the operator tree.
The address consists of three parts:
$$\left\langle exec\_id, ~~~~ (sop\_id_1, s_1),
\cdots, (sop\_id_d, s_d), ~~~~ op\_id \right\rangle$$
where $exec\_id$ identifies the hosting executor of the operator;
$op\_id$ represents the ID of the operator;
$(sop\_id_1, s_1),\cdots, (sop\_id_d, s_d)$ denotes the chain of ancestor
scope operators ($sop\_id_k$) and the corresponding scope instance IDs ($s_k$).
Actually, $\langle s_1, \cdots, s_d\rangle$ is the scope tag of the operator.

To facilitate hierarchical scheduling, each scope operator maintains a directory
of its child operators as a prefix tree, using the addresses of child operators
as the keys and the pointers to these operators as the values. In the prefix tree,
the operators of a scope instance are naturally grouped together as they share
the same prefix in their addresses, and thus can be quickly located.

\noindent\textbf{Operator Creation and Termination.}
\gqs{} is event-driven, it dynamically creates operators on request.
More specifically, sending a message to a non-existing operator triggers the creation
of this operator and all its non-existing ancestor scope operators.
A scope operator provides a system-level API \textit{TerminateScope(scope\_instance\_id)}
to terminate a scope instance, i.e., all the operators in this scope instance.
Terminating a scope operator will terminate all managed scope instances in cascade.
Messages sent to terminated operators are ignored.
\gqs{} recycles objects used for operators through memory management to avoid
excessive memory allocations.

\subsection{Parallelizing Progress Tracking}
\label{sec:distributed_progress}

In the mechanism of hierarchical progress tracking in Section~\ref{sec:progress},
tracking inside a scope requires the ingress vertex to notify the egress vertex of
the total number of instantiated scope instances. When the ingress and egress
vertices of a scope are parallelized, a single ingress operator may not be aware of
all the scope instances in this scope.

To tackle this problem, in a branch scope, each ingress operator broadcasts to all the egress
operators the largest ID of scope instances it has instantiated, and each egress operator takes
the maximum among these IDs as the total number of scope instances to be tracked.
In the case of loop scopes, every time an ingress operator only sees the EOS messages but no
data message from a specific loop iteration, it broadcasts to all the egress operators
the ID of the corresponding scope instance.
If an egress operator receives a specific scope instance ID from all the ingress instances,
it can conclude that this ID is the number of scope instances in this scope.

To reduce the number of operators created for a scoped dataflow,
\gqs{} skips creating operators that only receive EOS messages in their entire lifetime.
Instead, it handles EOS messages sent to these operators in their parent scope operator.
EOS messages sent to non-existing operators are buffered in their parent scope operator.
If the operator is created later, these buffered EOS messages are inserted into their mailboxes.
Otherwise, the parent scope operator emits EOS messages on behalf of the non-existing child operator
after receiving all the corresponding EOS messages.

\subsection{Load Balancing}
\label{sec:load_balancing}

Many realistic graphs are often scale-free, which may lead to a skewed workload distribution
among different tablets. And this skewness changes dynamically, as the graph accessing patterns
of the incoming queries continuously change.
In graph queries, graph-accessing operations are often the most costly part during the
execution. To facilitate load balancing between executors,
we deliberately partition the graph into more tablets, and migrate tablets
together with their graph-accessing operators across executors.
Upon migrating a tablet between hosts, we do not migrate the operators at
execution (as graph traversal queries are usually short-lived), but only redirect
the incoming queries.


\section{Evaluations}
\label{sec:evaluation}

In this section, we evaluate the performance of \gqs{} in the following aspects:
\begin{itemize}

\item (\textbf{E1}) We study the overall performance of \gqs{} by comparing
single-query latency with state-of-the-art graph query engines (Section~\ref{subsec:expr-benchmark}).

\item (\textbf{E2}) We study the effects and overheads of scopes on query
performance, by comparing the scoped dataflow with the Timely dataflow model (Section~\ref{subsec:expr-scope}).

\item (\textbf{E3}) We study how well \gqs{} can scale up in a many-core
server and scale out in a distributed cluster (Section~\ref{subsec:expr-scalability}).

\item (\textbf{E4}) We study how well \gqs{} can enforce performance
isolation and load balancing (Section~\ref{subsec:concurrency}).

\end{itemize}


\subsection{Experiment Setup}
\label{subsec:expr-setup}

\noindent\textbf{Benchmarks.}
We use two benchmarks in the experiments:
the LDBC Social Network Benchmark~\cite{ldbc-official-ref, ldbc} and
the Complex Query(CQ) benchmark.

LDBC is a popular benchmark of graph traversal queries.
We selected $12$ queries ($IC_1$ - $IC_{12}$) from the $14$
\textit{Interactive Complex Read} queries
in the LDBC benchmark, and exclude all the \textit{Analysis} ($BI_1$-$BI_{25}$),
\textit{Short Read} ($IS_1$-$IS_7$),
and \textit{Update} queries ($IU_1$-$IU_8$). $IC_{13}$ and $IC_{14}$ are excluded as they both
have shortest-path subqueries, which are typical graph analytics queries.
We use two LDBC datasets with scale factor $1$ and $100$, denoted as LDBC-1 and LDBC-100.
Table~\ref{table:data} shows the statistics of these two datasets.
For each query on both datasets we use the LDBC
generator to generate $50$ parameters.
\begin{table}[t]
\centering
\footnotesize
\begin{tabular}{|c|c|c|c|}
\hline
Dataset & \# Vertices & \# Edges & CSV Size \\ \hline
LDBC-1 & $3,181,364$ & $17,299,165$ & $882$M     \\ \hline
LDBC-100 & $282,637,871$ & $1,777,459,239$ & $88$G      \\ \hline
\end{tabular}
\caption{Dataset Statistics}
\vspace{-15pt}
\label{table:data}
\end{table}

\begin{figure*}[!t]
\centering
\begin{tabular}{c}
\subfigure[]{
  \includegraphics[width=.9\textwidth]{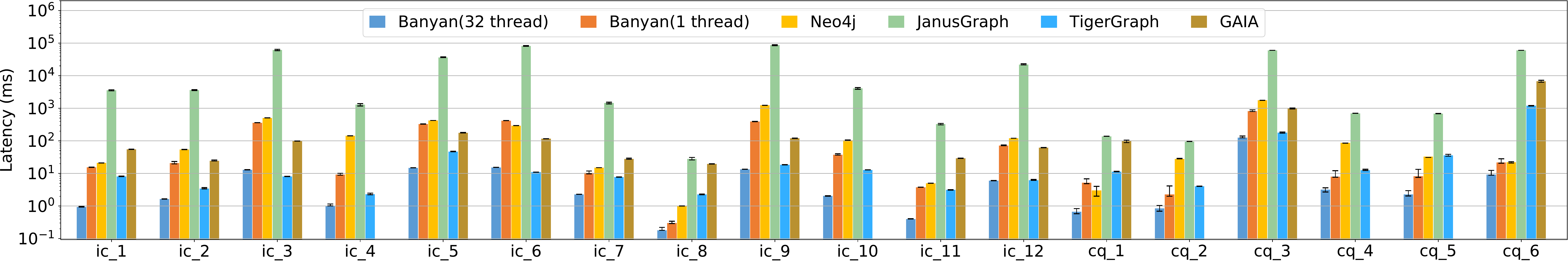}
  \label{fig:oversall-single}
}
\\
\subfigure[]{
  \includegraphics[width=.9\textwidth]{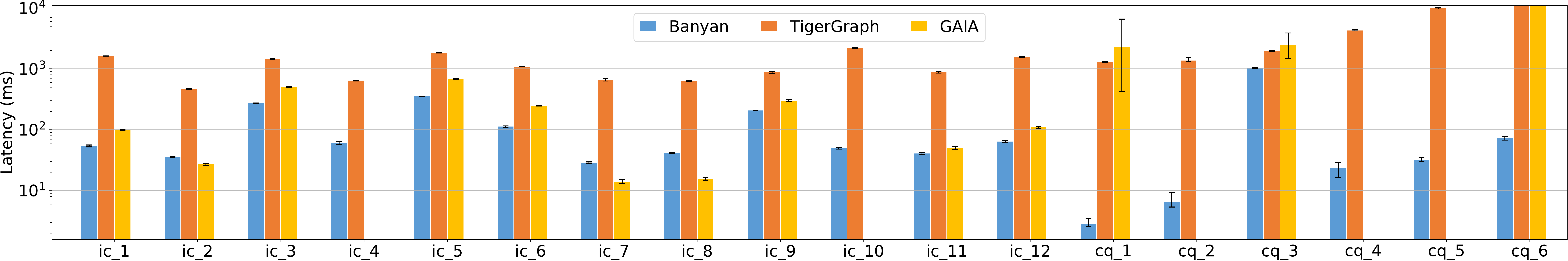}
  \label{fig:overall-distributed}
}
\end{tabular}
\vspace{-15pt}
\caption{(a) The single-query performance of \gqs{}, Neo4j, JanusGraph, TigerGraph and GAIA on a single machine with 32 cores. (b) The single-query performance of \gqs{}, TigerGraph and GAIA in a 4-node cluster (each node has 8 cores). }
\end{figure*}

Real-world service scenarios
(e.g. search engines) often select the top-k results from a limited size of recalled
candidates to guarantee interactive response. This is different from the query patterns
in LDBC (all LDBC queries require sorting the entire results).
To better study the effects of scopes, we compose the CQ benchmark with $6$ queries ($CQ_1$ - $CQ_6$)
by adjusting the LDBC queries
(e.g., 
removing the \textit{sort} operator) based on the two LDBC datasets.
Each CQ query
has $10$ parameters generated by the LDBC benchmark for both datasets.
The CQ queries are presented in Appendix~\ref{appendix:cq}.

\noindent\textbf{System Configurations.}
All the experiments are conducted on a cluster (up to $8$ machines) where each machine has
$755$G memory and $2$ Intel Xeon Platinum 8269CY CPUs
(each with $26$ physical cores and $52$ hyper-threads).

We choose four baseline systems from the most popular or latest graph databases/engines:
two single-machine ones---Neo4j $4.1.1$~\cite{neo4j} and JanusGraph $0.5.0$~\cite{janusgraph},
as well as two distributed ones---TigerGraph $3.1.0$~\cite{tigergraph}
and GAIA~\cite{gaia}.
We also compare scoped dataflow with Timely dataflow~\cite{naiad}
to study the effects of scopes on query performance.
For a fair comparison, we implement queries in the Timely dataflow
model using \gqs{} with scopes turned off.

Unless explicitly explained, all the experiments are
conducted in a container which has $32$ cores and $700$G memory.
We configure a cache size (if available) large enough to store the
entire datasets.
We build the same set of indexes for all systems,
i.e., a primary index on vertex ID for each type of vertices.
In graph databases,
we execute queries without transactions or as read-only transactions to
minimize the transaction overhead.
Configurations of the baseline systems
are as follows:

\begin{itemize}
  \item \emph{Neo4j 4.1.1}. We use $32$ worker threads and turn on query cache.

  \item \emph{JanusGraph 0.5.0}. We use BerkeleyJE $7.5.11$~\cite{berkeleyje} as the
   storage.

  \item \emph{TigerGraph 3.1.0}. The  distributed query mode is used in the distributed
  experiments. 
  TigerGraph requires installing a query before execution, and the
  installation takes much more time (more than $1$ min on average)
  than query execution. We exclude the installation time in the reported results.

  \item \emph{GAIA}. GAIA is only experimented on a subset of the LDBC and CQ queries,
  as it does not support $IC_4$, $IC_{10}$, $CQ_2$, $CQ_4$ and $CQ_5$
  \footnote{The authors of GAIA confirmed that their compiler is still under development and cannot support some Gremlin operators
  like \textit{sideEffect} and \textit{store}.}.
  The number of workers is set to $32$ such that GAIA can  use all the cores.

  \item \emph{\gqs{}}. We use a C++ version of the backend storage used by JanusGraph
  (BerkeleyDB $18.1.32$~\cite{berkeleydb}), and directly import the databases exported
  from JanusGraph. For each dataset, \gqs{} randomly partitions the graph into $64$ tablets.
  We rely on the load balancing mechanism
  to uniformly distribute tablets on executors. We apply loop scope on
  all the \textit{repeat} subqueries and branch scope on \textit{where}
  subqueries whose branches can be early canceled.
  Unless otherwise specified, we use $32$
  executors for query execution.
\end{itemize}

\noindent\textbf{Experiment Methodology.}
To flexibly control query submission, we extend the standard LDBC client
to allow specifying the number of concurrent queries ($W$) a client can submit.
Unless explicitly specified in \textbf{E3} and \textbf{E4},
we use $W=1$ throughout the experiments.
As LDBC/CQ queries are templates, unless otherwise specified,
we follow the LDBC benchmark convention and for each query report the average latency of
all the generated parameters.
Throughout this section, for each data point we run the corresponding
experiment $10$ times to warm up the system, and collect the results
from the following $10$ runs. We report the minimum, maximum, and average
values of the $10$ results in the figures. Queries run longer than $60$
seconds are marked as timeout.

\begin{figure*}[!tbh]
\centering
 \begin{tabular}{ccccc}
 \hspace{-10pt}
 \subfigure[]{
   \includegraphics[height=0.15\textwidth]{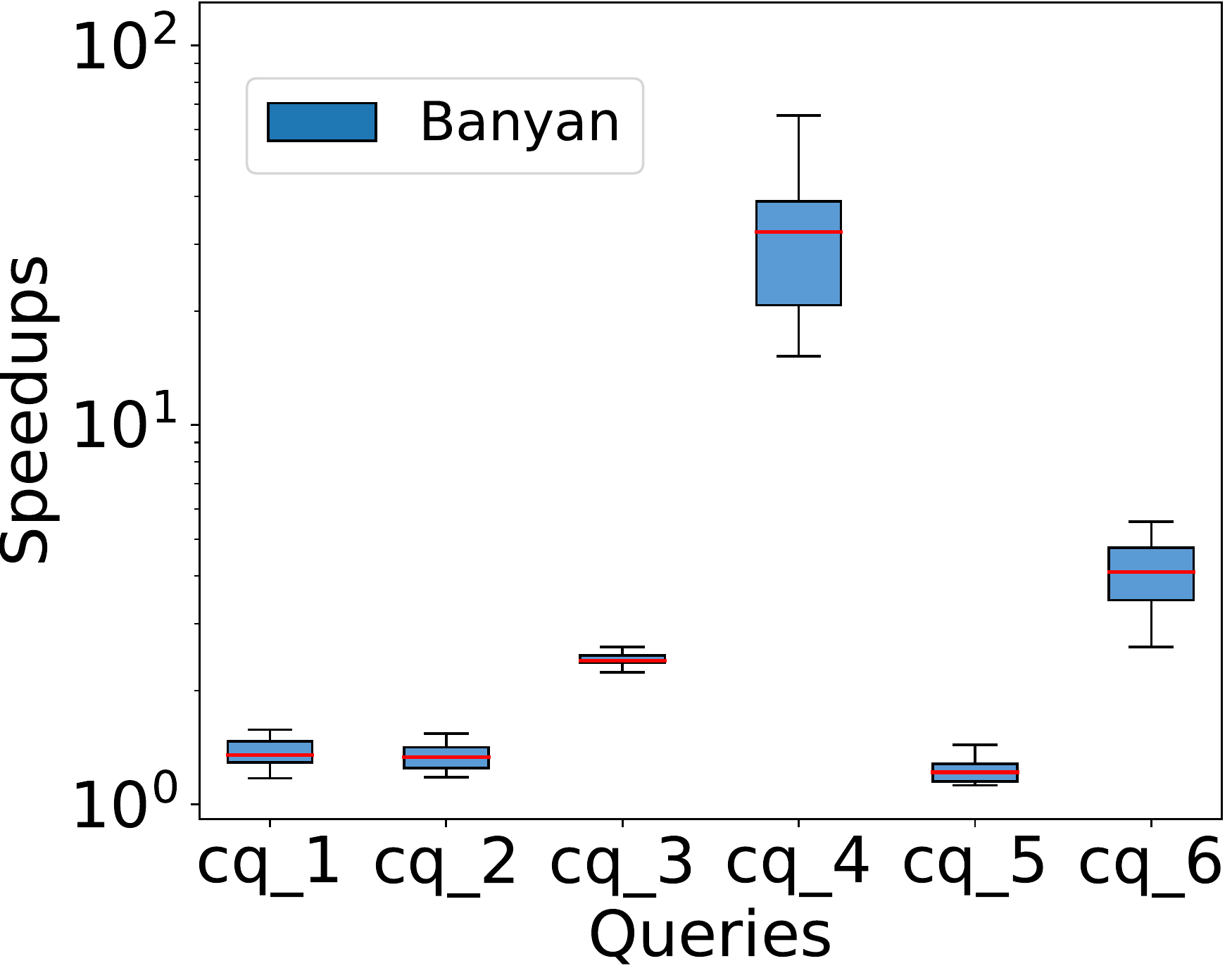}
   \label{fig:dataflow-model-comparison}
 }
 \hspace{-10pt}
 &
 \subfigure[]{
   \includegraphics[height=0.15\textwidth]{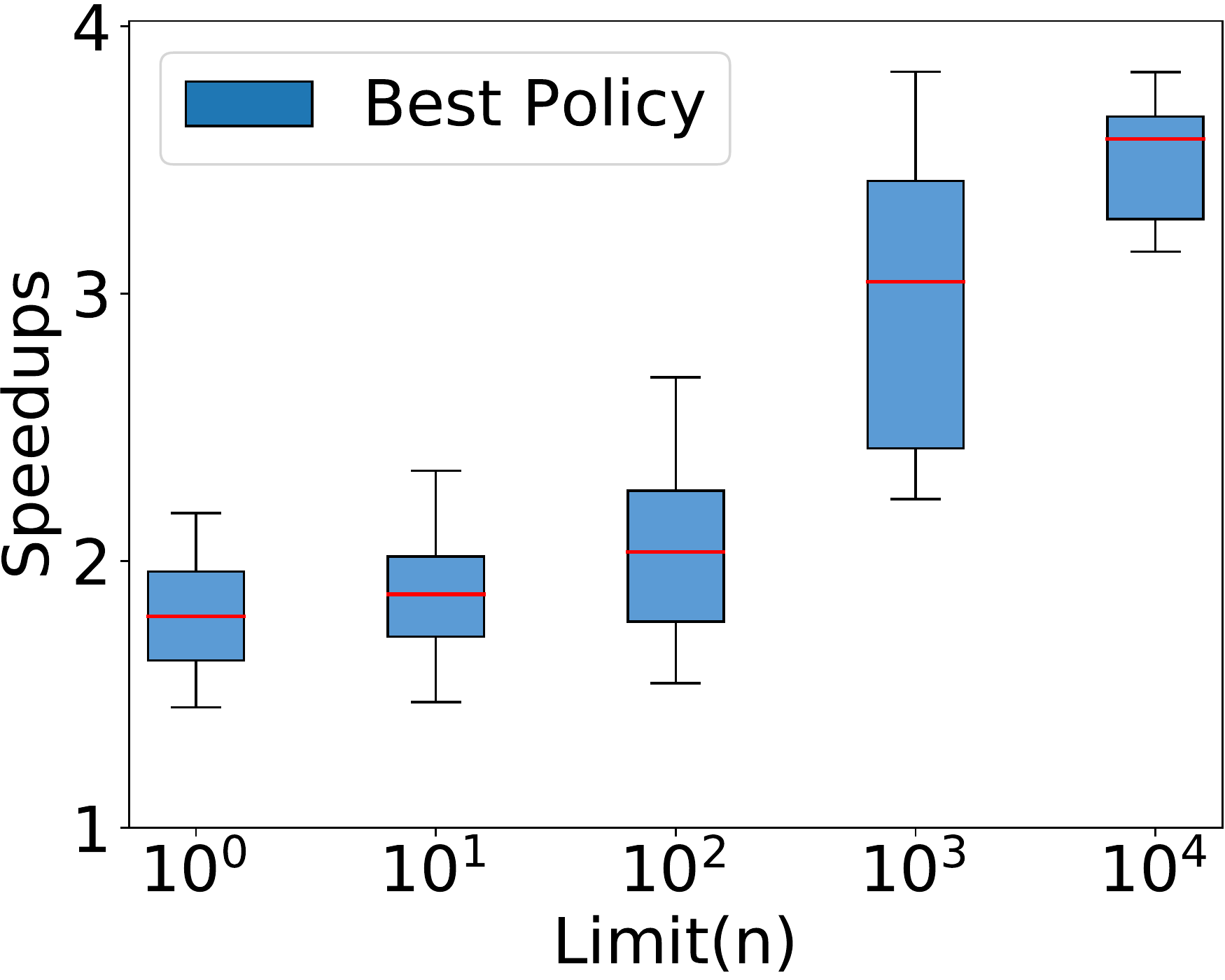}
   \label{fig:scheduling-policy-comparison}
 }
 \hspace{-10pt}
  &
 \subfigure[]{
   \includegraphics[height=0.15\textwidth]{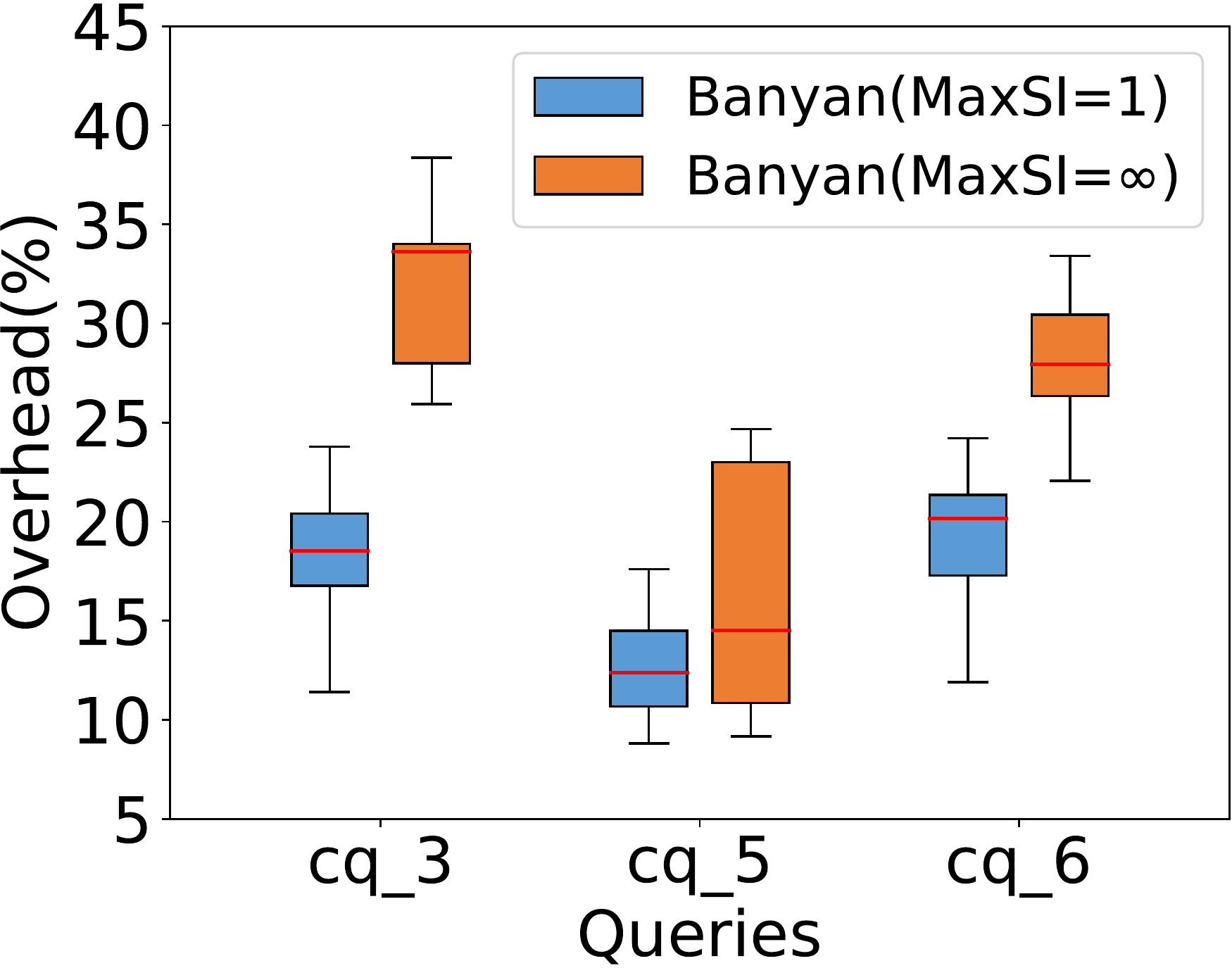}
   \label{fig:scope-overhead}
 }
  \hspace{-10pt}
  &
 \subfigure[]{
   \includegraphics[height=0.15\textwidth]{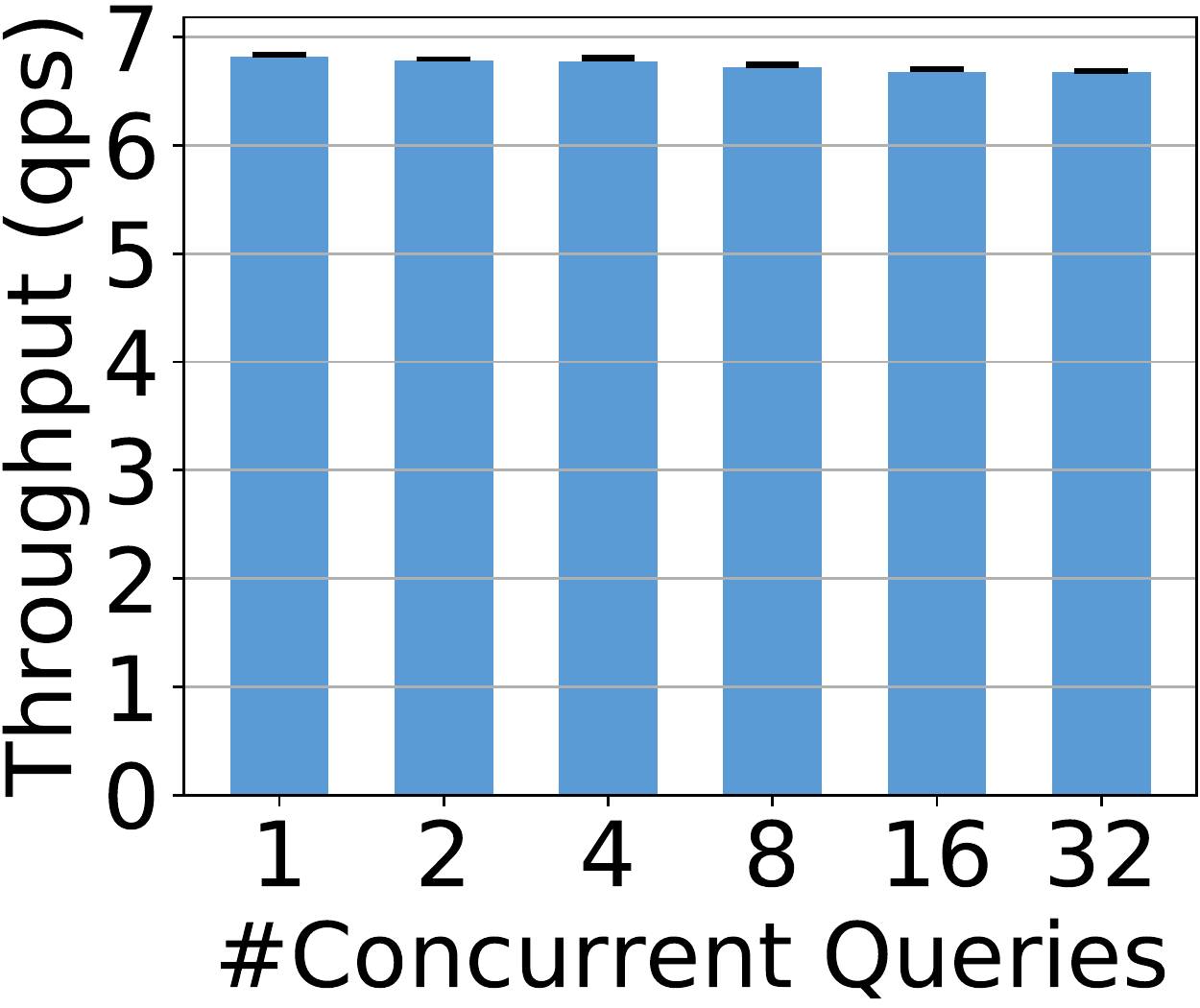}
   \label{fig:expr-cc-throughput}
 }
 \hspace{-10pt}
 &
 \subfigure[]{
   \includegraphics[height=0.15\textwidth]{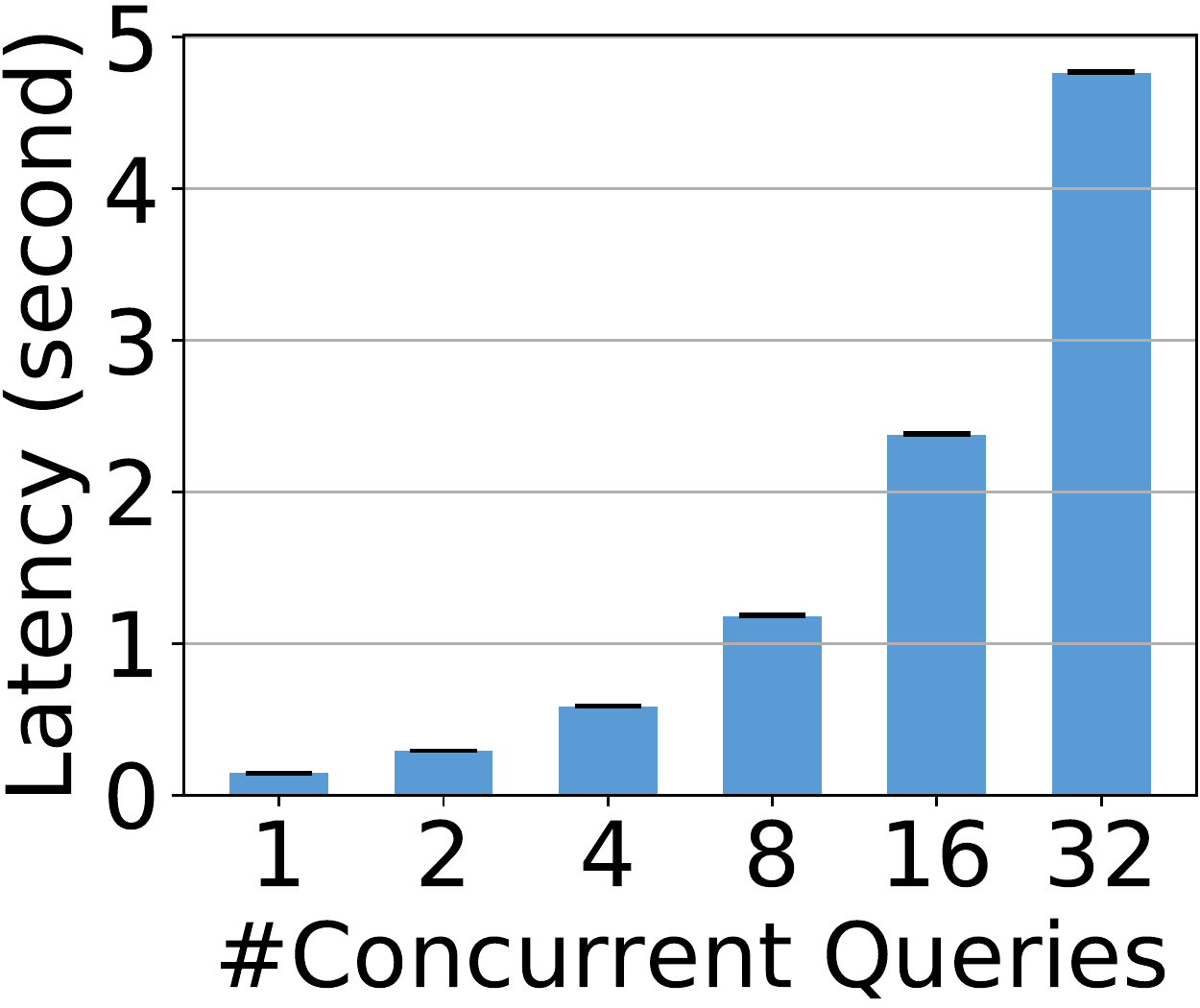}
   \label{fig:expr-cc-latency}
 }
 \end{tabular}
\vspace{-15pt}
\caption{(a) The per-parameter latency speedups of scoped dataflow compared to Timely.
(b) The per-parameter latency speedups of the \textit{best-case} scheduling policy
compared to the FIFO policy.
(c) The overheads of scope instantiation when MAX\_SI is set to 1 and unlimited.
The (d) throughput and (e) latency of \gqs{} under different numbers of concurrent queries.
}
\end{figure*}

\subsection{Overall System Performance}
\label{subsec:expr-benchmark}

In this section, we study the performance of \gqs{} by comparing its single-query
latencies on both the LDBC and CQ benchmarks with baseline systems.

\noindent\textbf{Single-machine.}
In this experiment, we use the LDBC-1 dataset so that baselines
like JanusGraph can finish most queries before timeout.
The results are depicted in Figure~\ref{fig:oversall-single}.

On the LDBC queries,
\gqs{} has $5$X to three orders of magnitude latency improvement over all
the baseline systems except for TigerGraph. \gqs{} outperforms TigerGraph
by up to $14$X for $10$ out of the $12$ LDBC queries (including
all the large queries), and is slightly slower on $IC_3$ and $IC_6$ (both are small queries).
As TigerGraph is not open-sourced, we cannot analyze how the query
installation helps in speeding up query execution.
This demonstrates that \gqs{} can well utilize the many-core parallelism.

On the CQ queries which can benefit more from the scoped dataflow,
the advantage of \gqs{} further widens, e.g., up to $130$X and $726$X faster than
TigerGraph and GAIA, respectively.
This improvement is led by the fine-grained control and
scheduling enabled by scoped dataflow: (1) subquery traversals
(\textit{where} subqueries in $CQ_3$, $CQ_4$, $CQ_5$ and $CQ_6$) can be early canceled,
and (2) customized scope-level scheduling policies (e.g., DFS in the loop of
$CQ_1$ and BFS in the loop of $CQ_2$) can trigger the (sub-)query cancellation earlier.
Note that GAIA’s support for traversal-level early cancellation is less efficient.
GAIA mixes messages with different `contexts’ in the same execution pipeline.
Canceling a specific traversal requires filtering all messages with the corresponding
context at each operator inside a scope. This inefficiency is especially reflected
by the performance of GAIA on $CQ_1$ and $CQ_6$. Differently, Banyan can directly drop
scope instances along with all related messages. Besides, GAIA cannot configure
scope-level scheduling policies, which influences its performance on $CQ_3$ and $CQ_6$.

Note that JanusGraph and Neo4j cannot parallelize queries whose traversal starts from a
single vertex. To make a fair comparison, we also present the single-thread
performance of \gqs{}, which outperforms both systems on all queries.

\noindent\textbf{Distributed.}
For the distributed experiment,
we use the LDBC-100 dataset and a cluster of $4$ worker nodes each hosting a
container of $8$ cores.
We compare only the distributed baseline systems.
The results are presented in Figure~\ref{fig:overall-distributed}.
\gqs{} outperforms TigerGraph by $5$X to $40$X on the LDBC benchmark,
and $2$X to three orders of magnitude on the
CQ benchmark. These results are consistent with the single-machine
experiment, as the benefits of efficiently utilizing hardware parallelism
and the scoped dataflow are still tenable in the distributed environment.
\gqs{} is faster than  GAIA on most of the LDBC queries and all the CQ queries,
except for three small queries ($IC_2$, $IC_7$ and $IC_8$).
This is because the graph partitioning in \gqs{}
can incur some overhead due to message passing between executors.

\subsection{Benchmark on Scoped Dataflow}
\label{subsec:expr-scope}

In this section, we use the CQ benchmark
on the LDBC-100 dataset to study:
(1) the effects of scoped dataflow, (2) the effects of scheduling
policy and (3) the overhead of scope instantiation.
We run each query with ten different parameters,
compute the speedup(overhead) between different competitors on each parameter,
and report the boxplot of all the per-parameter speedups(overheads) for each query.

\noindent\textbf{Effects of Scoped Dataflow.}
We report the boxplots of speedups brought by scopes
by comparing scoped dataflow against Timely in Figure~\ref{fig:dataflow-model-comparison}.
Note that, the Timely versions of queries are implemented
using \gqs{} with scopes turned off to make a fair comparison.
We can see the effects of scoped dataflow are:
\begin{itemize}
\item Query-dependent. E.g., the scoped dataflow brings $1.3$X to
$36$X latency improvement compared to Timely on average.
In particular, the speedup of scoped dataflow
on $CQ_1$ is relatively small, as $CQ_1$ has no subquery
which can be early canceled, and the speedup comes from
customizing the scope-level scheduling policy, i.e.,
using DFS to prioritize the loop iteration which generates the final
outputs. On the other hand, as $CQ_4$ contains a \textit{where} subquery nested
with a loop subquery, canceling a \textit{where} traversal can save a huge amount
of computation.
\item Data-dependent. The speedup of scoped dataflow varies on
different parameters, e.g., $14$X to $86$X on $CQ_4$.
This is because different traversals of the \textit{where} subquery in $CQ_4$
have very different computation cost.
\end{itemize}
Clearly, the optimal scoped dataflow plan should be determined by a query
compiler according to the query structure and data statistics.

\noindent\textbf{Effects of Scheduling Policy.}
In this experiment, we select $CQ_6$ (see Appendix ~\ref{appendix:cq})
and compare the latency of $CQ_6$ in \gqs{} between two cases:
(1) the \textit{best-policy} case where the intra-SI policy of the query and the \textit{where}
subquery are both set to DFS, and (2) the \textit{FIFO} case where all the
scheduling policies are set to FIFO in the query.
We also vary $n$ in $CQ_6$'s \textit{limit(n)} clause from $1$ to $10^4$.
Figure~\ref{fig:scheduling-policy-comparison}
shows the boxplots of speedups brought by \textit{best-policy} over \textit{FIFO}.
We can see that \textit{best-policy} outperforms \textit{FIFO} in all the settings.
By increasing the value of $n$ from $1$ to $10^4$, the speedups of \textit{best-policy}
widen from $1.8$X to $3.5$X. This is because \textit{FIFO} can result in more wasted
traversals which have no contribution to the final outputs, and this wastage becomes
worse with the increasing number of total traversals. Similar effects can be observed in
other CQ queries and thus omitted.
This experiment shows that customized scheduling policy is necessary for graph queries.

\noindent\textbf{Overhead of Scope Instantiation.}
To quantify the overhead of scope instantiation, we turn off early cancellation
in \gqs{}, use purely FIFO for scheduling, and compare its single query latencies with Timely.
We experiment on $CQ_3$, $CQ_5$ and $CQ_6$,
as these queries contains \textit{where} subqueries that can instantiate a large number of scope instances.
We remove the \textit{limit} clause in these queries to make sure that
both \gqs{} and Timely perform the same amount of traversals.
We report the boxplots of overheads caused by scope instantiation in
Figure~\ref{fig:scope-overhead}.

We can see that without limiting MAX\_SI, \gqs{} is on average $25\%$ slower than
Timely, as \gqs{} suffers from extra scheduling overheads among SIs and a high memory
pressure in this case. While, setting MAX\_SI in \gqs{} to $1$ shrinks the
performance differences to $13\%$.
Note that, MAX\_SI is an executor-local configuration. With $32$ executors
running in parallel, setting MAX\_SI to $1$
allows \gqs{} to instantiate at most $32$ concurrent SIs of a scope, which is
enough to saturate the multi-core parallelism. This experiment shows
that the overhead of scope instantiation is limited compared with
the benefits of scopes as shown in the first
experiment of \textbf{E2}.

\subsection{Scalability of \gqs{}}
\label{subsec:expr-scalability}

\begin{figure*}[!t]
\centering
\begin{tabular}{c}
\subfigure[]{
  \includegraphics[width=.92\textwidth]{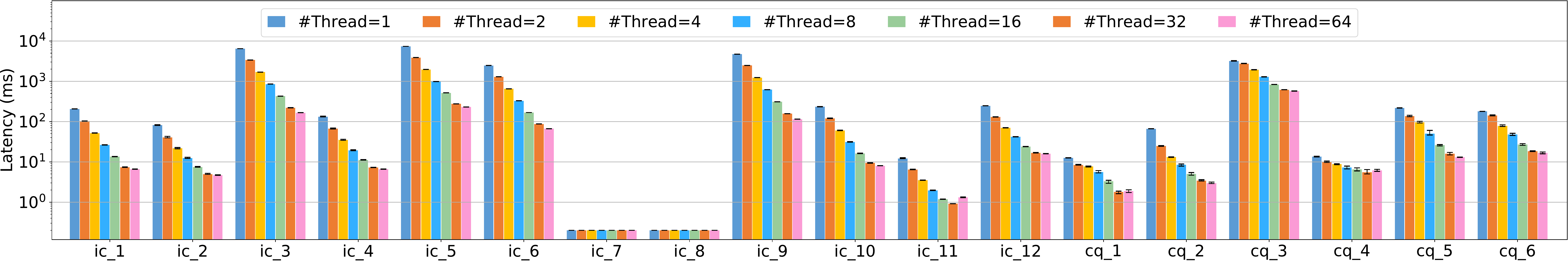}
  \label{fig:expr-threads}
}
\vspace{-5pt}
\\
\subfigure[]{
  \includegraphics[width=.92\textwidth]{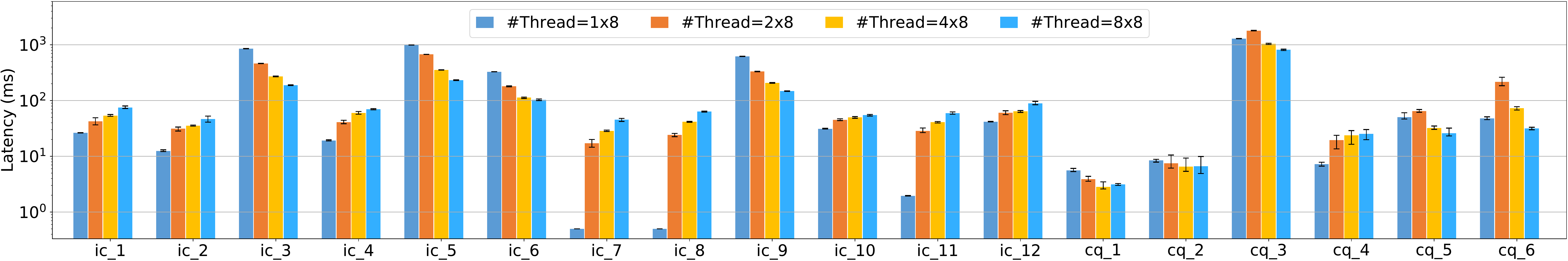}
  \label{fig:expr-cluster}
}
\end{tabular}
\vspace{-15pt}
\caption{(a) The scale-up performance of \gqs{} with increasing number of executors
         on a single machine. (b) The scale-out performance of \gqs{} with increasing number of worker nodes in the cluster.}
\end{figure*}

\begin{figure*}[!tbh]
\centering
 \begin{tabular}{cccc}
 \hspace{-10pt}
 \subfigure[]{
   \includegraphics[height=0.15\textwidth]{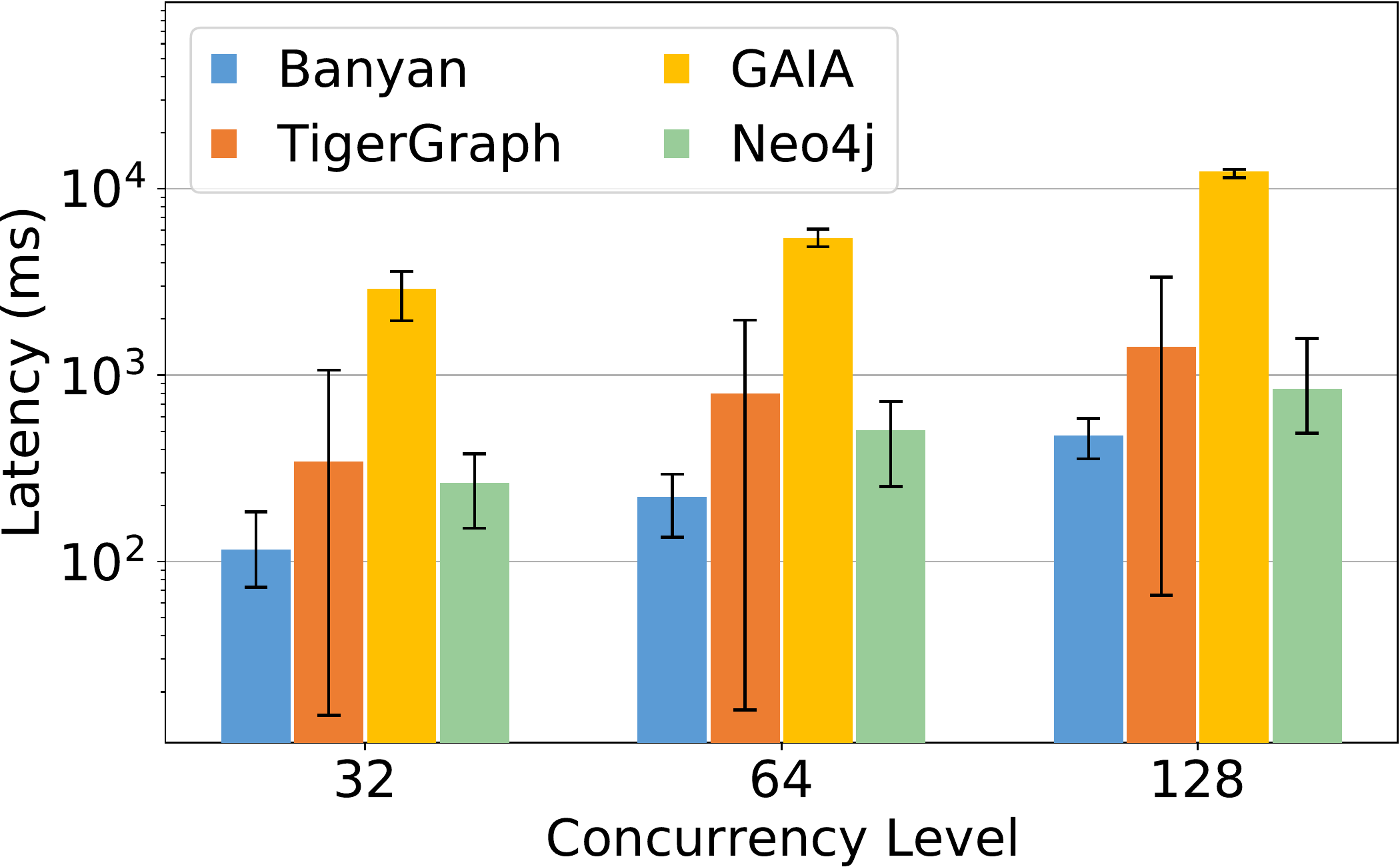}
 \label{fig:expr-service-sm}
 }
 \hspace{-10pt}
 &
 \subfigure[]{
   \includegraphics[height=0.15\textwidth]{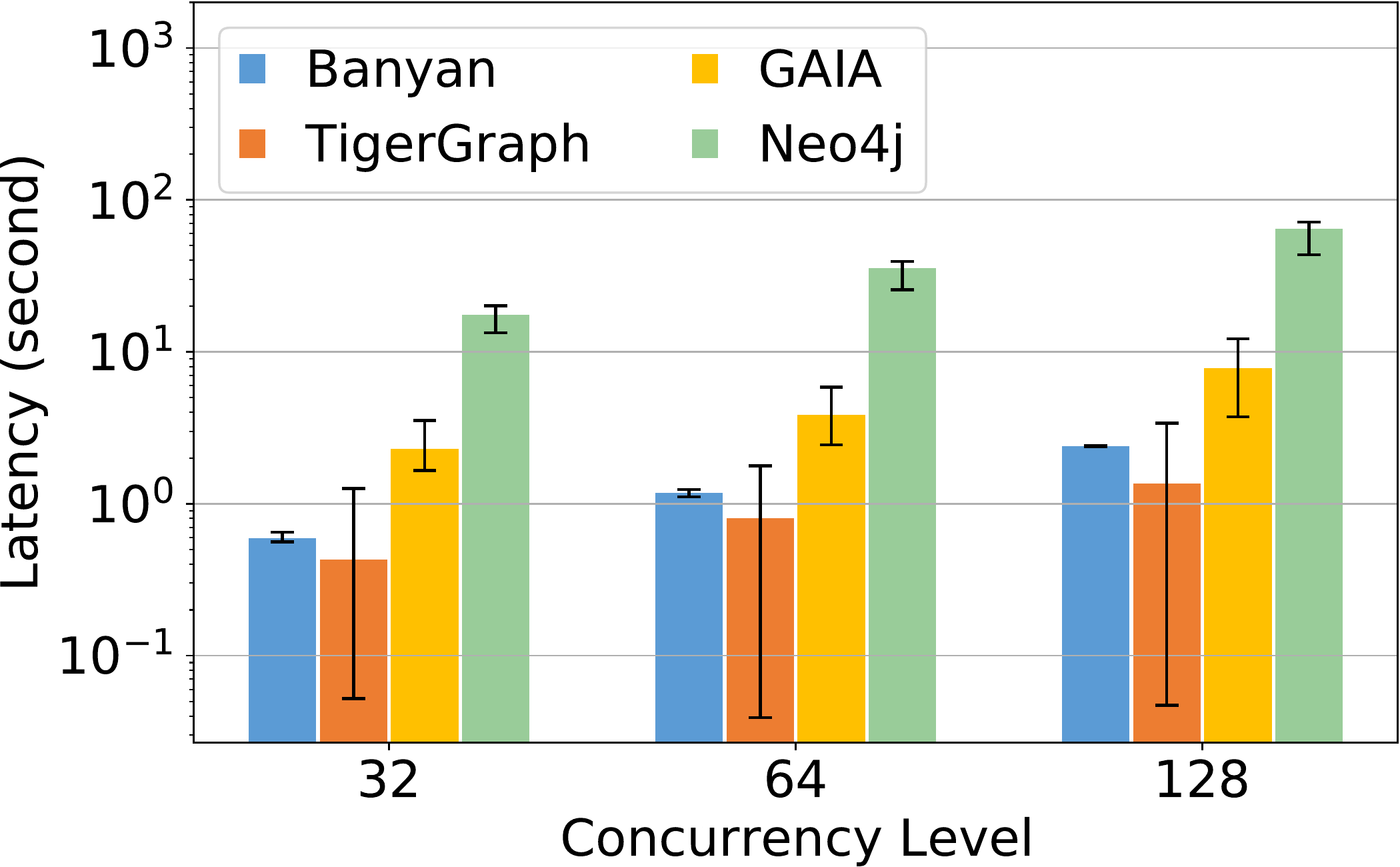}
 \label{fig:expr-service-big}
 }
 \hspace{-10pt}
 &
 \subfigure[]{
   \includegraphics[height=0.15\textwidth]{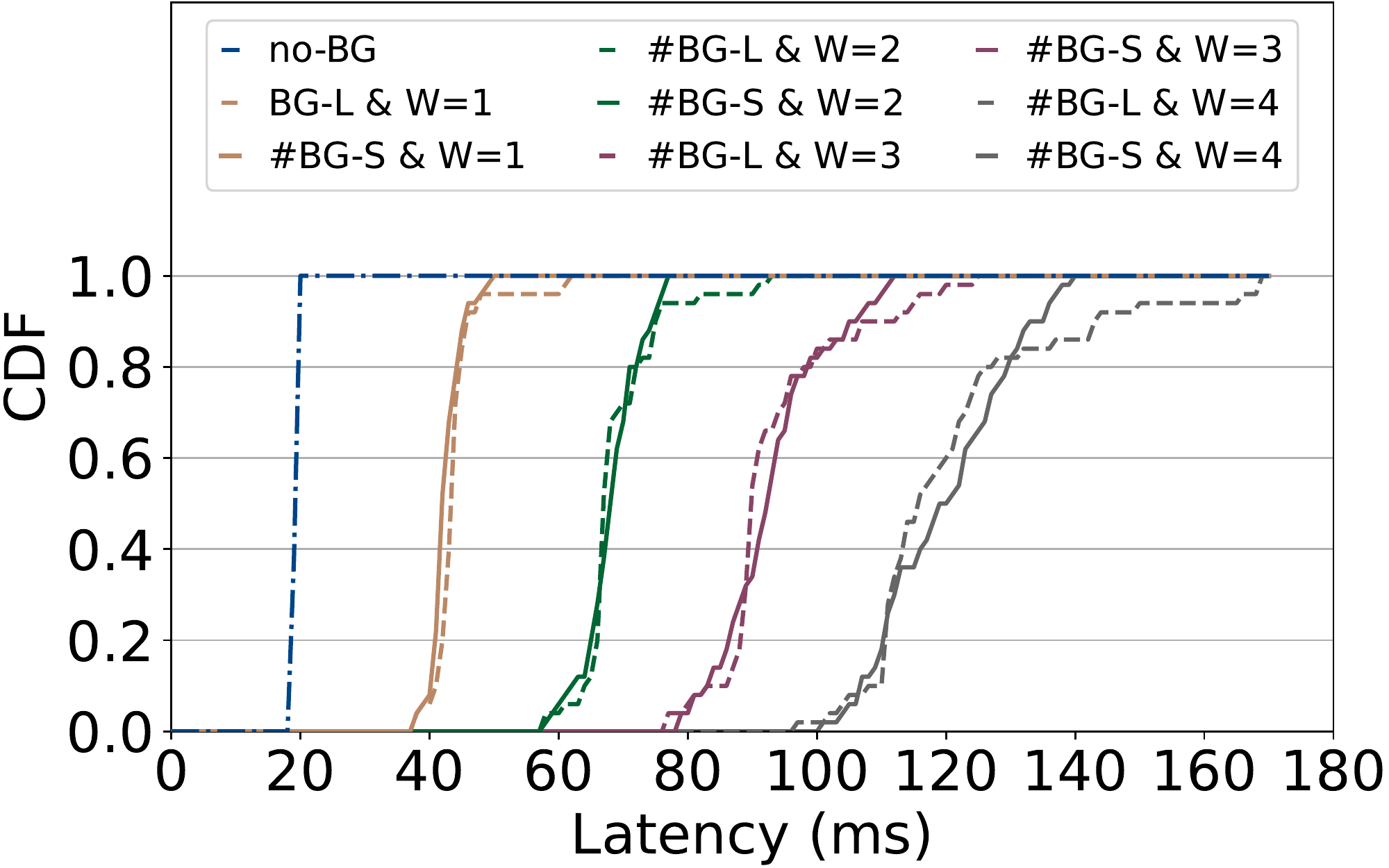}
   \label{fig:multi-tenant}
 }
 \hspace{-20pt}
 &
 \subfigure[]{
   \includegraphics[height=0.15\textwidth]{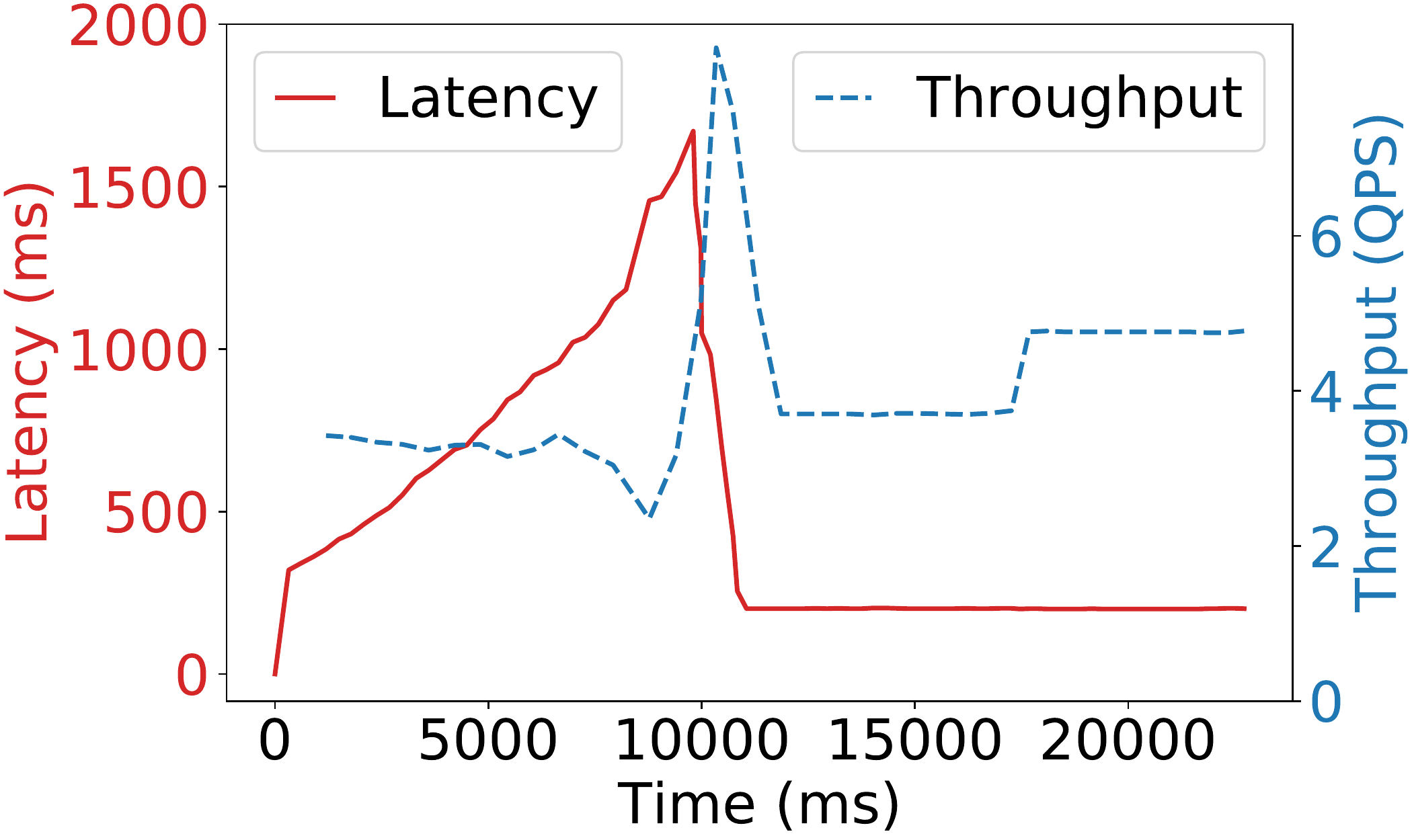}
   \label{fig:balancing}
 }
 \end{tabular}
\vspace{-15pt}
\caption{The latency of \gqs{}, TigerGraph, GAIA and Neo4j under different concurrent workloads:
(a) the latency of small queries and (b) the latency of large queries.
(c) The latency CDF of the foreground query under different background workloads.
(d) The throughput and latency of \gqs{} before and after load balancing.}
\end{figure*}

Next we conduct experiments to evaluate the scalability of Banyan from
three perspectives:
(1) the scale-up performance on a many-core machine,
(2) the scale-out performance in a cluster,
and (3) the throughput and latency when scaling the number of concurrent queries.
We use both the CQ and LDBC benchmarks with the LDBC-100 dataset in this experiment.

\noindent\textbf{Scale-Up.}
In this experiment, we use a single container and vary the number of cores from $1$ to $64$.
We report the query latency of \gqs{} in
Figure~\ref{fig:expr-threads}. We can see that the query
latency scales almost linearly up to $32$ cores.
This is because the architecture of \gqs{} can efficiently parallelize
a query dataflow into fine-grained operators and evenly distribute them across executors.
The performance improvement of \gqs{} stagnates when scaling up from 32
cores to 64. This is because per-executor workload becomes too small to
fully utilize allocated computation resources. On the other
hand, the hyper-threading has a negative impact on the cache locality when the
$64$ logical cores share $52$ physical cores.

\noindent\textbf{Scale-Out.}
In Figure~\ref{fig:expr-cluster}, we study the scale-out performance of \gqs{}.
We use a container of $8$ cores as the worker node,
and increase the number of worker nodes from $1$ to $8$.
We can observe that by increasing the number of worker nodes,
the latencies of large queries (e.g., $IC_3$, $IC_5$, $IC_6$ and $IC_9$)
decrease in a nearly linear manner. This shows that the scoped dataflow can be
well parallelized in a distributed environment, saturating the hardware parallelism.
For small queries (e.g., $IC_1$, $IC_2$, $IC_{10}$, $IC_{11}$ and $CQ_4$) with
limited computation to distribute, increasing the number of worker nodes
results in slightly worse query latency due to the network communication cost.
Note that we randomly
partition the graph data in \gqs{}, which is not optimized to reduce
inter-machine communication. As graph partitioning is orthogonal to our work,
we believe adopting advanced graph partitioning~\cite{partition}
can further improve the scale-out performance of \gqs{}.

\noindent\textbf{Scalability with Concurrent Queries.}
In this experiment, we use $IC_6$ with a fixed parameter to isolate the latency difference caused
by different queries and parameters. We vary the submission concurrency $W$ of
the client from $1$ to $32$, and report the throughput and latency of
\gqs{} in \Figure~\ref{fig:expr-cc-throughput} and~\ref{fig:expr-cc-latency}.
As shown in \Figure~\ref{fig:expr-cc-throughput}, by increasing the number of
concurrent queries, \gqs{} can provide a stable throughput (less than 2\%
throughput decrease when $W=32$). The query latency is increasing linearly
with more concurrent queries executing in the system
(\Figure~\ref{fig:expr-cc-latency}), which clearly shows that \gqs{} can fairly
time slice the CPU time among concurrent queries,
but incur little overhead when running more queries at the same time.
We can observe similar results on other queries.

\subsection{Performance Isolation \& Load Balancing}
\label{subsec:concurrency}

In this subsection, we study whether \gqs{} can enforce performance isolation
and load balancing. In the following experiments we use a fixed parameter for
each selected query as the effects of performance isolation and load balancing
are query/parameter-independent.

\noindent\textbf{Performance Isolation.}
We conduct two experiments to study performance isolation in \gqs{}.

In the first experiment, we compare the performance isolation in \gqs{}, Neo4j,
TigerGraph and GAIA using the LDBC benchmark with dataset LDBC-1.
In Figure~\ref{fig:expr-service-sm} and~\ref{fig:expr-service-big} we simulate a
mixed workload of large queries (using $IC_9$) and small queries (using $IC_1$),
and vary the submission concurrency $W$ from $32$ to $128$.

For different $W$, \gqs{} provides on average $2$X-$23$X better query latencies for small queries
compared with all the baselines (Figure~\ref{fig:expr-service-sm}),
and $3$X-$30$X performance boost on large queries compared with
GAIA and Neo4j (Figure~\ref{fig:expr-service-big}). TigerGraph runs
$50\%$ faster than $\gqs{}$ on the large queries, but sacrifices
its performance on small queries (on average $3.3$X worse than \gqs{}).
\gqs{} also achieves a much more stable latency performance than TigerGraph.
In Figure~\ref{fig:expr-service-sm}, when $W$ is set to $128$, the latency of
TigerGraph fluctuates on small queries at a range of nearly $3000$ms
($2$X of its average). In comparison,
the latency of \gqs{} only fluctuates within $50$ms ($10\%$ of its average).
This shows that the hierarchical scheduling in scoped dataflow
can guarantee performance isolation between queries.

In the second experiment, we conduct a controlled experiment
on how background workload impacts a foreground query.
We generate a background workload of different types of queries using the LDBC-100 dataset,
i.e., of purely large queries ($IC_9$) or purely small queries($IC_{10}$),
and vary the concurrency $W$ from $1$ to $4$.
While the background queries are executing,
we submit a small foreground query ($IC_1$) and collect its latency
using another client with $W=1$. For each background configuration,
we submit the foreground query for $100$ times
and plot the CDF of the query latency in Figure~\ref{fig:multi-tenant}.
We also plot the foreground latency without any background workload (No-BG)
as a baseline reference. We can see that when $W$ is fixed, the latency of
the foreground query is quite stable.
Changing the background workload from small queries (BG-S) to large queries (BG-L)
has negligible impact on the latency of the foreground query.
E.g., the $95\%$ percentile latency of the foreground query is only increased by
$3.5\%$ (resp. $9.5\%$) for $W=1$ (resp. $W=4$).
This experiment shows that \gqs{} can enforce performance isolation
so that large queries will not block small queries even when executed concurrently.

\noindent\textbf{Load Balancing.}
We simulate a scenario with skewed workload
to evaluate \gqs{}'s load balancing mechanism.
In specific, we distribute the $64$ tablets of LDBC-100
dataset among $8$ executors in a skewed manner:
evenly distributing $48$ tablets on $4$ executors
and the rest on the others. We repeatedly submit $IC_6$ every $270$ms.
At $t_1$ (around $9,000$ms), we re-balance the
distribution of tablets such that each executor has $8$ tablets.
At $t_2$ (around $17,000$ms), we set the submission interval to $210$ms.
The latency and throughput are reported in Figure~\ref{fig:balancing}.
We can see the query latency continuously increases when the workload is skewed.
After the re-balance ($t_1$), the query latency immediately drops,
and restores to a stable level (around $200$ms) after $3000$ms.
The throughput first bursts as \gqs{} is clearing the buffered work.
After increasing the input rate at $t_2$, the throughput increases to
nearly $1.3X$ of the initial throughput, while the latency remains stable.

\section{Related Work}
\label{sec:related}

\noindent\textbf{Graph Databases and Graph Engines}
Neo4j~\cite{neo4j}, Neptune~\cite{neptune}, TinkerPop~\cite{tinkerpop} and
JanusGraph~\cite{janusgraph} are single-machine graph databases.
TinkerPop, JanusGraph and Neptune only utilize multiple threads
for inter-query parallelization~\cite{janusgraph-transactions, tinkerpop-threaded-transactions,
neptune-queuing}. Neo4j supports intra-query parallelization, but can only parallelize
top-level traversals starting from different vertices.
On the contrary, \gqs{} can parallelize fine-grained subquery-level traversals.
Distributed graph databases such as TigerGraph~\cite{tigergraph}, DGraph~\cite{dgraph} and OrientDB~\cite{orientdb} are
mainly optimized for OLTP queries with both read and write operations. Differently, \gqs{}
focuses on answering read-only graph traversal queries in a GQS.
Grasper~\cite{chen2019grasper} proposes the Expert model to support adaptive
operator parallelization.
However, Grasper uses the same set of experts for concurrent queries, and thus cannot support
fine-grained control and enforce
performance isolation both inside and across queries as Banyan does.
GAIA~\cite{gaia} introduces a virtual \textit{scope} abstraction to facilitate data
dependency tracking in graph queries, so that the Gremlin queries with nested subqueries can be
correctly parallelized. Note that their scope is a logical concept used in compilation
to annotate subquery traversals, and is different from our scoped dataflow.
GAIA compiles Gremlin queries into topo-static dataflows,
and thus suffer the limitations discussed in Section~\ref{sec:motivation}.
Horton~\cite{sarwat2012horton} and Horton+~\cite{sarwat2013horton+} focus on
static query plan optimization and are orthogonal to \gqs{}.


G-SPARQL~\cite{sakr2012g}, Trinity.RDF~\cite{zeng2013distributed} and
Wukong~\cite{wukong} target for SPARQL queries. G-SPARQL~\cite{sakr2012g} uses a hybrid query
execution engine that can push parts of the query plan into the relational database.
Trinity.RDF~\cite{zeng2013distributed} utilizes graph exploration to answer SPARQL queries.
Wukong~\cite{wukong} supports concurrent execution with sub-query support.
The subquery in Wukong~\cite{wukong} is generated \textit{statically} and
is of \textit{coarser} granularity than scope instances, and thus cannot support
goal \textbf{O1} (see Section~\ref{sec:motivation}) required by a GQS.

Graph processing systems like Pregel~\cite{malewicz2010pregel}, PowerGraph~\cite{graphlab,
gonzalez2012powergraph} and GraphX~\cite{gonzalez2014graphx} focus on graph analysis workload,
whereas \gqs{} focuses on interactive graph traversal queries.

\noindent\textbf{Dataflow Engines.}
Dataflow systems such as \cite{hadoop, spark,isard2007dryad} adopt
BSP paradigm, and do not support cycles in the dataflow.
Flink~\cite{flink-website} supports cyclic dataflow but requires barrier synchronization
between loop iterations. Naiad~\cite{naiad} proposes
the Timely dataflow model for iterative and incremental computations, and cannot support
branch scopes for \textit{where} subqueries as \gqs{} does.
All the above dataflows are topo-static dataflow models.
Cilk~\cite{blumofe1995cilk} and CIEL~\cite{murray2011ciel} support
dynamic modification of the dataflow, but at the level of coarse-grained tasks.
This design may incur huge overhead to control nimble tasks like scope instances in scoped
dataflow. Whiz~\cite{grandl2021whiz}, Dandelion~\cite{Dandelion} and Optimus~\cite{Optimus}
can modify execution plans dynamically, whereas their target workloads have no requirement on
nimble, hierarchical management of computational resources as a GQS does.
Compared to existing dataflow models, the key difference of the scoped dataflow model is the
introduction of the scope construct. A scope can dynamically replicate its containing dataflow subgraph
at run time, creating a separate execution pipeline for different input data. This mechanism allows
concurrent execution and independent control on the replicated execution pipelines.
The scope construct
also provides a new way to support loops in dataflow, allowing users to control the
scheduling policy in cyclic computation.

Quickstep~\cite{quickstep} and Morsel~\cite{morsel} decompose a query into some
fine-grained tasks, and schedule tasks among multiple cores using a centralized flat task scheduler.
While \gqs{} adopts a similar parallelization mechanism for graph-accessing operators,
we adopt a novel hierarchical scheduling framework to recursively schedule operators
by their parent scope operators.
This allows \gqs{} to support performance isolation and customized scheduling policy
in nimble granularities.

\noindent\textbf{Actor-based Systems.}
Actor-based frameoworks~\cite{orleans-website, akka, bykov2011} use actors as the basic
computation primitives to build distributed and concurrent systems.
Orleans~\cite{orleans-website,bykov2011} provides a virtual actor mechanism where actors
are automatically instantiated when receiving a message, and reclaimed by the system when
being unused. The mechanism of operator activation in \gqs{} is similar to the virtual
actor mechanism in Orleans. The difference is that the lifetime of
an operator in \gqs is determined by the dataflow semantics, and managed by its parent scope operator.
Parent actors in Akka~\cite{akka} manage the creation and
termination of the actors they spawned. Different from Akka,
the scope operator in \gqs{} is also a scheduler for its child operators.
Furthermore, these systems mainly focus on the low-level programming framework,
but do not have a high-level dataflow computation model as \gqs{} does.

\section{Conclusions and Future Work}
We present a novel scoped dataflow model, and a new engine named
\gqs{} built on top of it for GQS. Scoped dataflow is targeted at
solving the need for sophisticated fine-grained control and scheduling in order
to fulfill stringent query latency and performance isolation
in a GQS. We demonstrate through \gqs{} that the new dataflow model
can be efficiently parallelized,
showing its scale-up ability on modern many-core architectures
and scale-out ability in a cluster.
The comparison with the state-of-the-art graph query engines shows
\gqs{} can provide at most $3$ orders of magnitude query latency improvement,
very stable system throughput and performance isolation.

Besides graph queries, the scoped dataflow model can also be applied to general service scenarios
that involve complex processing pipelines and require delicate control
of the pipeline execution.
As the forthcoming work, we plan to generalize the scoped dataflow model
and provide a high-level declarative language on top to help
users compose processing pipelines.

\appendix

\section{Complex Queries}
\label{appendix:cq}

\begin{complex_query}
\noindent
\begin{lstlisting}[style=QueryStyle, upquote=false]
g.V(person_id)
 .repeat(__.out(`knows')).times(5)
 .dedup().limit(n)
\end{lstlisting}
\label{cq:1}
\end{complex_query}

\noindent$CQ_1$: Given a start Person, find Persons that the start
Person is connected to by exactly 5 steps via the \textit{knows} relationship.
Return n distinct Person IDs.

\begin{complex_query}
\noindent
\begin{lstlisting}[style=QueryStyle, upquote=false]
g.V(person_id)
 .sideEffect(out(`workAt')
 .store(`companies'))
 .repeat(__.out(`knows'))
    .times(5)
    .emit(__.out(`workAt')
            .where(within(`companies'))
            .count().is(gt(0)))
 .dedup().limit(n)
\end{lstlisting}
\label{cq:2}
\end{complex_query}

\noindent$CQ_2$:
Given a start Person, find Persons that the start Peson is connected to by at most 5 steps via
the \textit{knows} relationship. Only Persons that \textit{workAt} the same Company with the start Person
are considered. Return n distinct Person IDs.

\begin{complex_query}
\noindent
\begin{lstlisting}[style=QueryStyle, upquote=false]
g.V(person_id)
 .out(`knows').union(identity(), out(`knows'))
 .dedup()
 .where(__.in(`hasCreator').out(`hasTag')
          .out(`hasType').values(`name')
          .filter{it.get().contains(`Country')})
 .order().by().limit(n)
\end{lstlisting}
\label{cq:3}
\end{complex_query}

\noindent$CQ_3$:
Given a start Person, find Persons that are his/her friends and friends of friends.
Only consider Persons that have created Messages with an attached Tag of TagClass `Country'.
Sort the Persons by their IDs and return the top-n Person IDs.

\begin{complex_query}
\noindent
\begin{lstlisting}[style=QueryStyle, upquote=false]
g.V(person_id)
 .sideEffect(out(`workAt')
 .store(`companies'))
 .out(`knows')
 .where(__.repeat(__.out(`knows'))
              .times(4)
              .emit(__.out(`workAt')
                      .where(within(`companies'))
                      .count().is(gt(0)))
          .dedup().count().is(gt(0)))
 .limit(n)
\end{lstlisting}
\label{cq:4}
\end{complex_query}

\noindent$CQ_4$:
Given a start Person, find Persons that are his/her friends. Only Persons that meet the following constraints are considered: for each S\_Person in Persons, find S\_Persons that S\_Person is connected to by at most 4 steps via the \textit{knows} relationship; If any Person in S\_Persons
\textit{workAt} the same Company as the start Person, S\_Person is selected as a candidate result.
Return n distinct Person IDs.

\begin{complex_query}
\noindent
\begin{lstlisting}[style=QueryStyle, upquote=false]
g.V(person_id)
 .sideEffect(out(`workAt')
 .store(`companies'))
 .repeat(__.out(`knows'))
    .times(5)
    .emit(__.out(`workAt')
            .where(within(`companies'))
            .count().is(gt(0)))
 .dedup()
 .where(__.in(`hasCreator').out(`hasTag')
          .out(`hasType').values(`name')
          .filter{it.get().contains(`Country')})
 .limit(n)
\end{lstlisting}
\label{cq:5}
\end{complex_query}

\noindent$CQ_5$:
Given a start Person, find Persons that the start Person is connected to by at most 5 steps via the
\textit{knows} relationship and \textit{workAt} the same Company with the start Person. Only consider
Persons that have created Messages with an attached Tag of TagClass `Country'. Return n distinct Person IDs.

\begin{complex_query}
\noindent
\begin{lstlisting}[style=QueryStyle, upquote=false]
g.V(person_id)
 .repeat(__.out(`knows')
           .where(__.in(`hasCreator')
                    .out(`hasTag').out(`hasType')
                    .values(`name')
                    .filter{it.get()
                    	.contains(`Country')}))
 .times(5).dedup().limit(n)
\end{lstlisting}
\label{cq:6}
\end{complex_query}

\noindent$CQ_6$:
Given a start Person, find Persons that the start Person is connected to by exactly 5 steps via the \textit{knows} relationship. Only Persons that meet the following constraints are considered: for each S\_Person in Persons, if every I\_Person in the path from the start Person to
S\_Person has created Messages with an attached Tag of TagClass `Country', S\_Person is selected as a candidate result. Return n distinct Person IDs.

\balance

\bibliographystyle{ACM-Reference-Format}
\bibliography{GQS}


\begin{thebibliography}{50}


\ifx \showCODEN    \undefined \def \showCODEN     #1{\unskip}     \fi
\ifx \showDOI      \undefined \def \showDOI       #1{#1}\fi
\ifx \showISBNx    \undefined \def \showISBNx     #1{\unskip}     \fi
\ifx \showISBNxiii \undefined \def \showISBNxiii  #1{\unskip}     \fi
\ifx \showISSN     \undefined \def \showISSN      #1{\unskip}     \fi
\ifx \showLCCN     \undefined \def \showLCCN      #1{\unskip}     \fi
\ifx \shownote     \undefined \def \shownote      #1{#1}          \fi
\ifx \showarticletitle \undefined \def \showarticletitle #1{#1}   \fi
\ifx \showURL      \undefined \def \showURL       {\relax}        \fi
\providecommand\bibfield[2]{#2}
\providecommand\bibinfo[2]{#2}
\providecommand\natexlab[1]{#1}
\providecommand\showeprint[2][]{arXiv:#2}

\bibitem[\protect\citeauthoryear{Akka}{Akka}{2021}]%
        {akka}
Akka \bibinfo{year}{2021}\natexlab{}.
\newblock \bibinfo{title}{Akka}.
\newblock \bibinfo{howpublished}{\url{http://akka.io//}}.
\newblock


\bibitem[\protect\citeauthoryear{Angles, Antal, et~al\mbox{.}}{Angles
  et~al\mbox{.}}{2020}]%
        {ldbc-official-ref}
\bibfield{author}{\bibinfo{person}{Renzo Angles},
  \bibinfo{person}{J{\'{a}}nos~Benjamin Antal}, {et~al\mbox{.}}}
  \bibinfo{year}{2020}\natexlab{}.
\newblock \showarticletitle{The {LDBC} {S}ocial {N}etwork {B}enchmark}.
\newblock \bibinfo{journal}{\emph{CoRR}}  \bibinfo{volume}{abs/2001.02299}
  (\bibinfo{year}{2020}).
\newblock
\showeprint[arxiv]{2001.02299}
\urldef\tempurl%
\url{http://arxiv.org/abs/2001.02299}
\showURL{%
\tempurl}


\bibitem[\protect\citeauthoryear{Apache Flink}{Apache Flink}{2021}]%
        {flink-website}
Apache Flink \bibinfo{year}{2021}\natexlab{}.
\newblock \bibinfo{title}{Apache Flink}.
\newblock \bibinfo{howpublished}{\url{https://flink.apache.org/}}.
\newblock


\bibitem[\protect\citeauthoryear{Apache Hadoop}{Apache Hadoop}{2021}]%
        {hadoop}
Apache Hadoop \bibinfo{year}{2021}\natexlab{}.
\newblock \bibinfo{title}{Apache Hadoop}.
\newblock \bibinfo{howpublished}{\url{https://github.com/apache/hadoop}}.
\newblock


\bibitem[\protect\citeauthoryear{Apache Spark}{Apache Spark}{2021}]%
        {spark}
Apache Spark \bibinfo{year}{2021}\natexlab{}.
\newblock \bibinfo{title}{Apache Spark}.
\newblock \bibinfo{howpublished}{\url{https://spark.apache.org/}}.
\newblock


\bibitem[\protect\citeauthoryear{Blumofe, Joerg, Kuszmaul, Leiserson, Randall,
  and Zhou}{Blumofe et~al\mbox{.}}{1995}]%
        {blumofe1995cilk}
\bibfield{author}{\bibinfo{person}{Robert~D Blumofe},
  \bibinfo{person}{Christopher~F Joerg}, \bibinfo{person}{Bradley~C Kuszmaul},
  \bibinfo{person}{Charles~E Leiserson}, \bibinfo{person}{Keith~H Randall},
  {and} \bibinfo{person}{Yuli Zhou}.} \bibinfo{year}{1995}\natexlab{}.
\newblock \showarticletitle{Cilk: An efficient multithreaded runtime system}.
\newblock \bibinfo{journal}{\emph{ACM SigPlan Notices}} \bibinfo{volume}{30},
  \bibinfo{number}{8} (\bibinfo{year}{1995}), \bibinfo{pages}{207--216}.
\newblock


\bibitem[\protect\citeauthoryear{Bykov, Geller, Kliot, Larus, Pandya, and
  Thelin}{Bykov et~al\mbox{.}}{2011}]%
        {bykov2011}
\bibfield{author}{\bibinfo{person}{Sergey Bykov}, \bibinfo{person}{Alan
  Geller}, \bibinfo{person}{Gabriel Kliot}, \bibinfo{person}{James~R Larus},
  \bibinfo{person}{Ravi Pandya}, {and} \bibinfo{person}{Jorgen Thelin}.}
  \bibinfo{year}{2011}\natexlab{}.
\newblock \showarticletitle{Orleans: cloud computing for everyone}. In
  \bibinfo{booktitle}{\emph{Proceedings of the 2nd ACM Symposium on Cloud
  Computing}}. ACM, \bibinfo{pages}{16}.
\newblock


\bibitem[\protect\citeauthoryear{Chandy and Lamport}{Chandy and
  Lamport}{1985}]%
        {chandy}
\bibfield{author}{\bibinfo{person}{K.~Mani Chandy} {and}
  \bibinfo{person}{Leslie Lamport}.} \bibinfo{year}{1985}\natexlab{}.
\newblock \showarticletitle{Distributed Snapshots: Determining Global States of
  Distributed Systems}.
\newblock \bibinfo{journal}{\emph{ACM Trans. Comput. Syst.}}
  (\bibinfo{year}{1985}).
\newblock


\bibitem[\protect\citeauthoryear{Chen, Li, Fang, Huang, Cheng, Zhang, Hou, and
  Yan}{Chen et~al\mbox{.}}{2019}]%
        {chen2019grasper}
\bibfield{author}{\bibinfo{person}{Hongzhi Chen}, \bibinfo{person}{Changji Li},
  \bibinfo{person}{Juncheng Fang}, \bibinfo{person}{Chenghuan Huang},
  \bibinfo{person}{James Cheng}, \bibinfo{person}{Jian Zhang},
  \bibinfo{person}{Yifan Hou}, {and} \bibinfo{person}{Xiao Yan}.}
  \bibinfo{year}{2019}\natexlab{}.
\newblock \showarticletitle{Grasper: A High Performance Distributed System for
  OLAP on Property Graphs}. In \bibinfo{booktitle}{\emph{Proceedings of the ACM
  Symposium on Cloud Computing}}. \bibinfo{pages}{87--100}.
\newblock


\bibitem[\protect\citeauthoryear{Cypher}{Cypher}{2021}]%
        {cypher}
Cypher \bibinfo{year}{2021}\natexlab{}.
\newblock \bibinfo{booktitle}{\emph{Cypher Query Language}}.
\newblock
\urldef\tempurl%
\url{https://neo4j.com/developer/cypher/}
\showURL{%
Retrieved Dec 20, 2021 from \tempurl}


\bibitem[\protect\citeauthoryear{DGraph}{DGraph}{2021}]%
        {dgraph}
DGraph \bibinfo{year}{2021}\natexlab{}.
\newblock \bibinfo{title}{DGraph: Not everything can fit in rows and columns}.
\newblock \bibinfo{howpublished}{\url{https://dgraph.io//}}.
\newblock


\bibitem[\protect\citeauthoryear{Gonzalez, Low, Gu, Bickson, and
  Guestrin}{Gonzalez et~al\mbox{.}}{2012}]%
        {gonzalez2012powergraph}
\bibfield{author}{\bibinfo{person}{Joseph~E Gonzalez}, \bibinfo{person}{Yucheng
  Low}, \bibinfo{person}{Haijie Gu}, \bibinfo{person}{Danny Bickson}, {and}
  \bibinfo{person}{Carlos Guestrin}.} \bibinfo{year}{2012}\natexlab{}.
\newblock \showarticletitle{Powergraph: Distributed graph-parallel computation
  on natural graphs}. In \bibinfo{booktitle}{\emph{Presented as part of the
  10th {USENIX} Symposium on Operating Systems Design and Implementation
  ({OSDI}'12)}}. \bibinfo{pages}{17--30}.
\newblock


\bibitem[\protect\citeauthoryear{Gonzalez, Xin, Dave, Crankshaw, Franklin, and
  Stoica}{Gonzalez et~al\mbox{.}}{2014}]%
        {gonzalez2014graphx}
\bibfield{author}{\bibinfo{person}{Joseph~E Gonzalez},
  \bibinfo{person}{Reynold~S Xin}, \bibinfo{person}{Ankur Dave},
  \bibinfo{person}{Daniel Crankshaw}, \bibinfo{person}{Michael~J Franklin},
  {and} \bibinfo{person}{Ion Stoica}.} \bibinfo{year}{2014}\natexlab{}.
\newblock \showarticletitle{Graphx: Graph processing in a distributed dataflow
  framework}. In \bibinfo{booktitle}{\emph{11th {USENIX} Symposium on Operating
  Systems Design and Implementation ({OSDI}'14)}}. \bibinfo{pages}{599--613}.
\newblock


\bibitem[\protect\citeauthoryear{Grandl, Singhvi, Viswanathan, and
  Akella}{Grandl et~al\mbox{.}}{2021}]%
        {grandl2021whiz}
\bibfield{author}{\bibinfo{person}{Robert Grandl}, \bibinfo{person}{Arjun
  Singhvi}, \bibinfo{person}{Raajay Viswanathan}, {and} \bibinfo{person}{Aditya
  Akella}.} \bibinfo{year}{2021}\natexlab{}.
\newblock \showarticletitle{Whiz: Data-Driven Analytics Execution}. In
  \bibinfo{booktitle}{\emph{18th $\{$USENIX$\}$ Symposium on Networked Systems
  Design and Implementation ($\{$NSDI$\}$ 21)}}. \bibinfo{pages}{407--423}.
\newblock


\bibitem[\protect\citeauthoryear{GraphDB Annual Growth}{GraphDB Annual
  Growth}{[n.d.]}]%
        {graphdb-annual-growth}
GraphDB Annual Growth \bibinfo{year}{[n.d.]}\natexlab{}.
\newblock \bibinfo{title}{Graph Database Market Size, Share and Global Market
  Forecast to 2024}.
\newblock
  \bibinfo{howpublished}{\url{https://www.marketsandmarkets.com/Market-Reports/graph-database-market-126230231.html}}.
\newblock


\bibitem[\protect\citeauthoryear{Hendrickson and Kolda}{Hendrickson and
  Kolda}{2000}]%
        {partition}
\bibfield{author}{\bibinfo{person}{Bruce Hendrickson} {and}
  \bibinfo{person}{Tamara~G Kolda}.} \bibinfo{year}{2000}\natexlab{}.
\newblock \showarticletitle{Graph partitioning models for parallel computing}.
\newblock \bibinfo{journal}{\emph{Parallel computing}} \bibinfo{volume}{26},
  \bibinfo{number}{12} (\bibinfo{year}{2000}), \bibinfo{pages}{1519--1534}.
\newblock


\bibitem[\protect\citeauthoryear{Isard, Budiu, Yu, Birrell, and Fetterly}{Isard
  et~al\mbox{.}}{2007}]%
        {isard2007dryad}
\bibfield{author}{\bibinfo{person}{Michael Isard}, \bibinfo{person}{Mihai
  Budiu}, \bibinfo{person}{Yuan Yu}, \bibinfo{person}{Andrew Birrell}, {and}
  \bibinfo{person}{Dennis Fetterly}.} \bibinfo{year}{2007}\natexlab{}.
\newblock \showarticletitle{Dryad: distributed data-parallel programs from
  sequential building blocks}. In \bibinfo{booktitle}{\emph{Proceedings of the
  2nd ACM SIGOPS/EuroSys European Conference on Computer Systems 2007}}.
  \bibinfo{pages}{59--72}.
\newblock


\bibitem[\protect\citeauthoryear{Janusgraph}{Janusgraph}{2021a}]%
        {janusgraph}
Janusgraph \bibinfo{year}{2021}\natexlab{a}.
\newblock \bibinfo{title}{JanusGraph: Distributed, open source, massively
  scalable graph database}.
\newblock \bibinfo{howpublished}{\url{https://janusgraph.org//}}.
\newblock


\bibitem[\protect\citeauthoryear{Janusgraph}{Janusgraph}{2021b}]%
        {janusgraph-transactions}
Janusgraph \bibinfo{year}{2021}\natexlab{b}.
\newblock \bibinfo{title}{JanusGraph Transactions}.
\newblock
  \bibinfo{howpublished}{\url{https://docs.janusgraph.org/basics/transactions/}}.
\newblock


\bibitem[\protect\citeauthoryear{Ke, Isard, and Yu}{Ke et~al\mbox{.}}{2013}]%
        {Optimus}
\bibfield{author}{\bibinfo{person}{Qifa Ke}, \bibinfo{person}{Michael Isard},
  {and} \bibinfo{person}{Yuan Yu}.} \bibinfo{year}{2013}\natexlab{}.
\newblock \showarticletitle{Optimus: A Dynamic Rewriting Framework for
  Data-Parallel Execution Plans}. In \bibinfo{booktitle}{\emph{Proceedings of
  the 8th ACM European Conference on Computer Systems}} (Prague, Czech
  Republic) \emph{(\bibinfo{series}{EuroSys '13})}.
  \bibinfo{publisher}{Association for Computing Machinery},
  \bibinfo{address}{New York, NY, USA}, \bibinfo{pages}{15–28}.
\newblock
\showISBNx{9781450319942}
\urldef\tempurl%
\url{https://doi.org/10.1145/2465351.2465354}
\showDOI{\tempurl}


\bibitem[\protect\citeauthoryear{LDBC Benchmark}{LDBC Benchmark}{2021}]%
        {ldbc}
LDBC Benchmark \bibinfo{year}{2021}\natexlab{}.
\newblock \bibinfo{title}{LDBC}.
\newblock \bibinfo{howpublished}{\url{http://ldbcouncil.org}}.
\newblock


\bibitem[\protect\citeauthoryear{Leis, Boncz, Kemper, and Neumann}{Leis
  et~al\mbox{.}}{2014}]%
        {morsel}
\bibfield{author}{\bibinfo{person}{Viktor Leis}, \bibinfo{person}{Peter Boncz},
  \bibinfo{person}{Alfons Kemper}, {and} \bibinfo{person}{Thomas Neumann}.}
  \bibinfo{year}{2014}\natexlab{}.
\newblock \showarticletitle{Morsel-Driven Parallelism: A NUMA-Aware Query
  Evaluation Framework for the Many-Core Age} \emph{(\bibinfo{series}{SIGMOD
  '14})}. \bibinfo{publisher}{Association for Computing Machinery},
  \bibinfo{address}{New York, NY, USA}, \bibinfo{pages}{743–754}.
\newblock
\showISBNx{9781450323765}


\bibitem[\protect\citeauthoryear{Low, Bickson, Gonzalez, Guestrin, Kyrola, and
  Hellerstein}{Low et~al\mbox{.}}{2012}]%
        {graphlab}
\bibfield{author}{\bibinfo{person}{Yucheng Low}, \bibinfo{person}{Danny
  Bickson}, \bibinfo{person}{Joseph Gonzalez}, \bibinfo{person}{Carlos
  Guestrin}, \bibinfo{person}{Aapo Kyrola}, {and} \bibinfo{person}{Joseph~M.
  Hellerstein}.} \bibinfo{year}{2012}\natexlab{}.
\newblock \showarticletitle{Distributed GraphLab: A Framework for Machine
  Learning and Data Mining in the Cloud}.
\newblock \bibinfo{journal}{\emph{Proc. VLDB Endow.}} \bibinfo{volume}{5},
  \bibinfo{number}{8} (\bibinfo{date}{April} \bibinfo{year}{2012}),
  \bibinfo{pages}{716–727}.
\newblock
\showISSN{2150-8097}
\urldef\tempurl%
\url{https://doi.org/10.14778/2212351.2212354}
\showDOI{\tempurl}


\bibitem[\protect\citeauthoryear{Malewicz, Austern, Bik, Dehnert, Horn, Leiser,
  and Czajkowski}{Malewicz et~al\mbox{.}}{2010}]%
        {malewicz2010pregel}
\bibfield{author}{\bibinfo{person}{Grzegorz Malewicz},
  \bibinfo{person}{Matthew~H Austern}, \bibinfo{person}{Aart~JC Bik},
  \bibinfo{person}{James~C Dehnert}, \bibinfo{person}{Ilan Horn},
  \bibinfo{person}{Naty Leiser}, {and} \bibinfo{person}{Grzegorz Czajkowski}.}
  \bibinfo{year}{2010}\natexlab{}.
\newblock \showarticletitle{Pregel: a system for large-scale graph processing}.
  In \bibinfo{booktitle}{\emph{Proceedings of the 2010 ACM SIGMOD International
  Conference on Management of data}}. \bibinfo{pages}{135--146}.
\newblock


\bibitem[\protect\citeauthoryear{Mellbin and {\AA}kerlund}{Mellbin and
  {\AA}kerlund}{2017}]%
        {Mellbin}
\bibfield{author}{\bibinfo{person}{Ragnar Mellbin} {and} \bibinfo{person}{Felix
  {\AA}kerlund}.} \bibinfo{year}{2017}\natexlab{}.
\newblock \showarticletitle{Multi-threaded execution of Cypher queries}.
\newblock
  \bibinfo{howpublished}{\url{https://lup.lub.lu.se/student-papers/record/8926892/file/8926896.pdf}}.
\newblock  (\bibinfo{year}{2017}).
\newblock
\newblock
\shownote{Technical Report.}


\bibitem[\protect\citeauthoryear{Murray, McSherry, Isaacs, Isard, Barham, and
  Abadi}{Murray et~al\mbox{.}}{2013}]%
        {naiad}
\bibfield{author}{\bibinfo{person}{Derek~G Murray}, \bibinfo{person}{Frank
  McSherry}, \bibinfo{person}{Rebecca Isaacs}, \bibinfo{person}{Michael Isard},
  \bibinfo{person}{Paul Barham}, {and} \bibinfo{person}{Mart{\'\i}n Abadi}.}
  \bibinfo{year}{2013}\natexlab{}.
\newblock \showarticletitle{Naiad: a timely dataflow system}. In
  \bibinfo{booktitle}{\emph{Proceedings of the Twenty-Fourth ACM Symposium on
  Operating Systems Principles}}. \bibinfo{pages}{439--455}.
\newblock


\bibitem[\protect\citeauthoryear{Murray, Schwarzkopf, Smowton, Smith,
  Madhavapeddy, and Hand}{Murray et~al\mbox{.}}{2011}]%
        {murray2011ciel}
\bibfield{author}{\bibinfo{person}{Derek~G Murray}, \bibinfo{person}{Malte
  Schwarzkopf}, \bibinfo{person}{Christopher Smowton}, \bibinfo{person}{Steven
  Smith}, \bibinfo{person}{Anil Madhavapeddy}, {and} \bibinfo{person}{Steven
  Hand}.} \bibinfo{year}{2011}\natexlab{}.
\newblock \showarticletitle{Ciel: a universal execution engine for distributed
  data-flow computing}. In \bibinfo{booktitle}{\emph{Proc. 8th ACM/USENIX
  Symposium on Networked Systems Design and Implementation}}.
  \bibinfo{pages}{113--126}.
\newblock


\bibitem[\protect\citeauthoryear{Neo4j}{Neo4j}{2021a}]%
        {neo4jplan}
Neo4j \bibinfo{year}{2021}\natexlab{a}.
\newblock \bibinfo{title}{Neo4j Execution Plans}.
\newblock
  \bibinfo{howpublished}{\url{https://neo4j.com/docs/developer-manual/3.0/cypher/execution-plans/}}.
\newblock


\bibitem[\protect\citeauthoryear{Neo4j}{Neo4j}{2021b}]%
        {neo4j}
Neo4j \bibinfo{year}{2021}\natexlab{b}.
\newblock \bibinfo{title}{Neo4j: The Fastest Path To Graph Success}.
\newblock \bibinfo{howpublished}{\url{https://neo4j.com//}}.
\newblock


\bibitem[\protect\citeauthoryear{Neptune}{Neptune}{2021a}]%
        {neptune}
Neptune \bibinfo{year}{2021}\natexlab{a}.
\newblock \bibinfo{title}{AWS Neptune}.
\newblock \bibinfo{howpublished}{\url{https://aws.amazon.com/neptune/}}.
\newblock


\bibitem[\protect\citeauthoryear{Neptune}{Neptune}{2021b}]%
        {neptune-queuing}
Neptune \bibinfo{year}{2021}\natexlab{b}.
\newblock \bibinfo{title}{Query queuing in Amazon Neptune}.
\newblock
  \bibinfo{howpublished}{\url{https://docs.aws.amazon.com/neptune/latest/userguide/access-graph-queuing.html}}.
\newblock


\bibitem[\protect\citeauthoryear{Oracle Berkeley DB}{Oracle Berkeley
  DB}{2017}]%
        {berkeleyje}
Oracle Berkeley DB \bibinfo{year}{2017}\natexlab{}.
\newblock \bibinfo{title}{BerkeleyJE 7.5.11}.
\newblock
\newblock
\urldef\tempurl%
\url{https://docs.oracle.com/cd/E17277_02/html/index.html}
\showURL{%
Retrieved Dec 20, 2021 from \tempurl}


\bibitem[\protect\citeauthoryear{Oracle Berkeley DB}{Oracle Berkeley
  DB}{2021}]%
        {berkeleydb}
Oracle Berkeley DB \bibinfo{year}{2021}\natexlab{}.
\newblock \bibinfo{title}{BerkeleyDB 18.1.32}.
\newblock
  \bibinfo{howpublished}{\url{https://www.oracle.com/database/technologies/related/berkeleydb-downloads.html}}.
\newblock


\bibitem[\protect\citeauthoryear{OrientDB}{OrientDB}{2021}]%
        {orientdb}
OrientDB \bibinfo{year}{2021}\natexlab{}.
\newblock \bibinfo{title}{OrientDB}.
\newblock \bibinfo{howpublished}{\url{https://www.orientdb.org}}.
\newblock


\bibitem[\protect\citeauthoryear{Orleans}{Orleans}{2021}]%
        {orleans-website}
Orleans \bibinfo{year}{2021}\natexlab{}.
\newblock \bibinfo{title}{Microsoft Orleans}.
\newblock \bibinfo{howpublished}{\url{https://dotnet.github.io/orleans/}}.
\newblock


\bibitem[\protect\citeauthoryear{Patel, Deshmukh, Zhu, Potti, Zhang, Spehlmann,
  Memisoglu, and Saurabh}{Patel et~al\mbox{.}}{2018}]%
        {quickstep}
\bibfield{author}{\bibinfo{person}{Jignesh~M. Patel}, \bibinfo{person}{Harshad
  Deshmukh}, \bibinfo{person}{Jianqiao Zhu}, \bibinfo{person}{Navneet Potti},
  \bibinfo{person}{Zuyu Zhang}, \bibinfo{person}{Marc Spehlmann},
  \bibinfo{person}{Hakan Memisoglu}, {and} \bibinfo{person}{Saket Saurabh}.}
  \bibinfo{year}{2018}\natexlab{}.
\newblock \showarticletitle{Quickstep: A Data Platform Based on the Scaling-up
  Approach}.
\newblock \bibinfo{journal}{\emph{Proc. VLDB Endow.}} \bibinfo{volume}{11},
  \bibinfo{number}{6} (\bibinfo{date}{Feb.} \bibinfo{year}{2018}),
  \bibinfo{pages}{663–676}.
\newblock
\showISSN{2150-8097}


\bibitem[\protect\citeauthoryear{Qian, Min, et~al\mbox{.}}{Qian
  et~al\mbox{.}}{2021}]%
        {gaia}
\bibfield{author}{\bibinfo{person}{Zhengping Qian}, \bibinfo{person}{Chenqiang
  Min}, {et~al\mbox{.}}} \bibinfo{year}{2021}\natexlab{}.
\newblock \showarticletitle{{GAIA}: A System for Interactive Analysis on
  Distributed Graphs Using a High-Level Language}. In
  \bibinfo{booktitle}{\emph{18th {USENIX} Symposium on Networked Systems Design
  and Implementation ({NSDI} 21)}}. \bibinfo{publisher}{{USENIX} Association}.
\newblock
\showISBNx{978-1-939133-21-2}


\bibitem[\protect\citeauthoryear{Rossbach, Yu, Currey, Martin, and
  Fetterly}{Rossbach et~al\mbox{.}}{2013}]%
        {Dandelion}
\bibfield{author}{\bibinfo{person}{Christopher~J. Rossbach},
  \bibinfo{person}{Yuan Yu}, \bibinfo{person}{Jon Currey},
  \bibinfo{person}{Jean-Philippe Martin}, {and} \bibinfo{person}{Dennis
  Fetterly}.} \bibinfo{year}{2013}\natexlab{}.
\newblock \showarticletitle{Dandelion: A Compiler and Runtime for Heterogeneous
  Systems}. In \bibinfo{booktitle}{\emph{Proceedings of the Twenty-Fourth ACM
  Symposium on Operating Systems Principles}} (Farminton, Pennsylvania)
  \emph{(\bibinfo{series}{SOSP '13})}. \bibinfo{publisher}{Association for
  Computing Machinery}, \bibinfo{address}{New York, NY, USA},
  \bibinfo{pages}{49–68}.
\newblock
\showISBNx{9781450323888}
\urldef\tempurl%
\url{https://doi.org/10.1145/2517349.2522715}
\showDOI{\tempurl}


\bibitem[\protect\citeauthoryear{Sakr, Elnikety, and He}{Sakr
  et~al\mbox{.}}{2012}]%
        {sakr2012g}
\bibfield{author}{\bibinfo{person}{Sherif Sakr}, \bibinfo{person}{Sameh
  Elnikety}, {and} \bibinfo{person}{Yuxiong He}.}
  \bibinfo{year}{2012}\natexlab{}.
\newblock \showarticletitle{G-SPARQL: a hybrid engine for querying large
  attributed graphs}. In \bibinfo{booktitle}{\emph{Proceedings of the 21st ACM
  international conference on Information and knowledge management}}. ACM,
  \bibinfo{pages}{335--344}.
\newblock


\bibitem[\protect\citeauthoryear{Sarwat, Elnikety, He, and Kliot}{Sarwat
  et~al\mbox{.}}{2012}]%
        {sarwat2012horton}
\bibfield{author}{\bibinfo{person}{Mohamed Sarwat}, \bibinfo{person}{Sameh
  Elnikety}, \bibinfo{person}{Yuxiong He}, {and} \bibinfo{person}{Gabriel
  Kliot}.} \bibinfo{year}{2012}\natexlab{}.
\newblock \showarticletitle{Horton: Online query execution engine for large
  distributed graphs}. In \bibinfo{booktitle}{\emph{2012 IEEE 28th
  International Conference on Data Engineering}}. IEEE,
  \bibinfo{pages}{1289--1292}.
\newblock


\bibitem[\protect\citeauthoryear{Sarwat, Elnikety, He, and Mokbel}{Sarwat
  et~al\mbox{.}}{2013}]%
        {sarwat2013horton+}
\bibfield{author}{\bibinfo{person}{Mohamed Sarwat}, \bibinfo{person}{Sameh
  Elnikety}, \bibinfo{person}{Yuxiong He}, {and} \bibinfo{person}{Mohamed~F
  Mokbel}.} \bibinfo{year}{2013}\natexlab{}.
\newblock \showarticletitle{Horton+: A distributed system for processing
  declarative reachability queries over partitioned graphs}.
\newblock \bibinfo{journal}{\emph{Proceedings of the VLDB Endowment}}
  \bibinfo{volume}{6}, \bibinfo{number}{14} (\bibinfo{year}{2013}),
  \bibinfo{pages}{1918--1929}.
\newblock


\bibitem[\protect\citeauthoryear{Shao, Wang, and Li}{Shao
  et~al\mbox{.}}{2013}]%
        {shao2013trinity}
\bibfield{author}{\bibinfo{person}{Bin Shao}, \bibinfo{person}{Haixun Wang},
  {and} \bibinfo{person}{Yatao Li}.} \bibinfo{year}{2013}\natexlab{}.
\newblock \showarticletitle{Trinity: A distributed graph engine on a memory
  cloud}. In \bibinfo{booktitle}{\emph{Proceedings of the 2013 ACM SIGMOD
  International Conference on Management of Data}}. ACM,
  \bibinfo{pages}{505--516}.
\newblock


\bibitem[\protect\citeauthoryear{Shi, Yao, Chen, Chen, and Li}{Shi
  et~al\mbox{.}}{2016}]%
        {wukong}
\bibfield{author}{\bibinfo{person}{Jiaxin Shi}, \bibinfo{person}{Youyang Yao},
  \bibinfo{person}{Rong Chen}, \bibinfo{person}{Haibo Chen}, {and}
  \bibinfo{person}{Feifei Li}.} \bibinfo{year}{2016}\natexlab{}.
\newblock \showarticletitle{Fast and concurrent $\{$RDF$\}$ queries with
  RDMA-based distributed graph exploration}. In \bibinfo{booktitle}{\emph{12th
  $\{$USENIX$\}$ Symposium on Operating Systems Design and Implementation
  ($\{$OSDI$\}$ 16)}}. \bibinfo{pages}{317--332}.
\newblock


\bibitem[\protect\citeauthoryear{TigerGraph}{TigerGraph}{2021}]%
        {tigergraph}
TigerGraph \bibinfo{year}{2021}\natexlab{}.
\newblock \bibinfo{title}{TigerGraph 3.1.0}.
\newblock \bibinfo{howpublished}{\url{https://www.tigergraph.com}}.
\newblock


\bibitem[\protect\citeauthoryear{Tinkerpop}{Tinkerpop}{2021}]%
        {tinkerpop}
Tinkerpop \bibinfo{year}{2021}\natexlab{}.
\newblock \bibinfo{title}{Apache {T}inkerpop}.
\newblock \bibinfo{howpublished}{\url{http://tinkerpop.apache.org/}}.
\newblock


\bibitem[\protect\citeauthoryear{Tinkerpopo}{Tinkerpopo}{2021}]%
        {tinkerpop-threaded-transactions}
Tinkerpopo \bibinfo{year}{2021}\natexlab{}.
\newblock \bibinfo{title}{Tinkerpopo Threaded Transactions}.
\newblock
\newblock
\urldef\tempurl%
\url{http://tinkerpop.apache.org/docs/current/reference/#_threaded_transactions}
\showURL{%
Retrieved Dec 20, 2021 from \tempurl}


\bibitem[\protect\citeauthoryear{Yan, Cheng, {\"O}zsu, Yang, Lu, Lui, Zhang,
  and Ng}{Yan et~al\mbox{.}}{2016}]%
        {yan2016general}
\bibfield{author}{\bibinfo{person}{Da Yan}, \bibinfo{person}{James Cheng},
  \bibinfo{person}{M~Tamer {\"O}zsu}, \bibinfo{person}{Fan Yang},
  \bibinfo{person}{Yi Lu}, \bibinfo{person}{John Lui}, \bibinfo{person}{Qizhen
  Zhang}, {and} \bibinfo{person}{Wilfred Ng}.} \bibinfo{year}{2016}\natexlab{}.
\newblock \showarticletitle{A general-purpose query-centric framework for
  querying big graphs}.
\newblock \bibinfo{journal}{\emph{Proceedings of the VLDB Endowment}}
  \bibinfo{volume}{9}, \bibinfo{number}{7} (\bibinfo{year}{2016}),
  \bibinfo{pages}{564--575}.
\newblock


\bibitem[\protect\citeauthoryear{Zeng, Yang, Wang, Shao, and Wang}{Zeng
  et~al\mbox{.}}{2013}]%
        {zeng2013distributed}
\bibfield{author}{\bibinfo{person}{Kai Zeng}, \bibinfo{person}{Jiacheng Yang},
  \bibinfo{person}{Haixun Wang}, \bibinfo{person}{Bin Shao}, {and}
  \bibinfo{person}{Zhongyuan Wang}.} \bibinfo{year}{2013}\natexlab{}.
\newblock \showarticletitle{A distributed graph engine for web scale RDF data}.
\newblock \bibinfo{journal}{\emph{Proceedings of the VLDB Endowment}}
  \bibinfo{volume}{6}, \bibinfo{number}{4} (\bibinfo{year}{2013}),
  \bibinfo{pages}{265--276}.
\newblock


\bibitem[\protect\citeauthoryear{Zhang, Chen, and Chen}{Zhang
  et~al\mbox{.}}{2015}]%
        {Zhang2015}
\bibfield{author}{\bibinfo{person}{Kaiyuan Zhang}, \bibinfo{person}{Rong Chen},
  {and} \bibinfo{person}{Haibo Chen}.} \bibinfo{year}{2015}\natexlab{}.
\newblock \showarticletitle{NUMA-Aware Graph-Structured Analytics}.
\newblock \bibinfo{journal}{\emph{SIGPLAN Not.}} (\bibinfo{year}{2015}).
\newblock


\bibitem[\protect\citeauthoryear{Zhao and Han}{Zhao and Han}{2010}]%
        {zhao2010graph}
\bibfield{author}{\bibinfo{person}{Peixiang Zhao} {and} \bibinfo{person}{Jiawei
  Han}.} \bibinfo{year}{2010}\natexlab{}.
\newblock \showarticletitle{On graph query optimization in large networks}.
\newblock \bibinfo{journal}{\emph{Proceedings of the VLDB Endowment}}
  \bibinfo{volume}{3}, \bibinfo{number}{1-2} (\bibinfo{year}{2010}),
  \bibinfo{pages}{340--351}.
\newblock


\end{thebibliography}

\end{document}